\documentclass[preprint]{aastex61}

\usepackage{amsmath}

\newcommand\pcc{\;{\rm cm}^{-3}}
\newcommand\Msun{\; M_{\odot}}
\newcommand\kms{\; {\rm km}\;{\rm s}^{-1}}

\newcommand\erg{\; {\rm erg}}
\newcommand\mh{\; m_{\rm H}}
\newcommand\cm{\;{\rm cm}}
\newcommand\yr{\; {\rm yr}}
\newcommand\Myr{\;{\rm Myr}}

\newcommand\kpc{\;{\rm kpc}}

\newcommand\sfrunit{\Msun \kpc^{-2} \yr^{-1}}

\newcommand\Kel{\;{\rm K}}

\newcommand\simgt{\lower.5ex\hbox{$\; \buildrel > \over \sim \;$}}
\newcommand\simlt{\lower.5ex\hbox{$\; \buildrel < \over \sim \;$}}

\newcommand{\vect}[1]{\mathbf{#1}}
\newcommand{\revise}[1]{#1}
\newcommand{\second}[1]{#1}

{}

\begin{document}


\title{Galactic Disk Winds Driven by Cosmic Ray Pressure}

\author[0000-0002-2491-8700]{S. Alwin Mao}
\author[0000-0002-0509-9113]{Eve C. Ostriker}
\affiliation{Department of Astrophysical Sciences, Princeton University, Princeton, NJ 08544, USA}
\shortauthors{Mao \& Ostriker}
\shorttitle{Winds Driven by Cosmic Rays}
\email{alwin@princeton.edu, eco@astro.princeton.edu}

\begin{abstract}
  Cosmic ray pressure gradients transfer energy and momentum to
  extraplanar gas in disk galaxies, potentially driving significant
  mass loss as galactic winds.  This may be particularly important for
  launching high-velocity outflows of ``cool'' ($T\lesssim 10^4\Kel$) gas.
  We study cosmic-ray driven disk winds using a simplified
  semi-analytic model assuming streamlines follow the large-scale
  gravitational potential gradient.  We consider scaled Milky Way-like
  potentials including a disk, bulge, and halo with a range of halo
  velocities $V_{H} = 50-300 \kms$, and streamline footpoints with
  radii in the disk
  $R_0=1-16\kpc$
\second{at height $1\kpc$}. Our solutions cover a wide range of footpoint gas
  velocity $u_{0}$, magnetic-to-cosmic-ray pressure ratio,
  gas-to-cosmic-ray pressure ratio, and angular momentum.  Cosmic ray
  streaming at the Alfv\'en speed enables the effective sound speed
  $C_{\rm eff}$ to \textit{increase} from the footpoint to a critical point
  where $C_{\rm eff,c} = u_c \sim V_{H}$; this differs from thermal
  winds in which $C_{\rm eff}$ \textit{decreases} outward.  The critical point
  is typically at a height of $1-6\kpc$ from the disk, increasing with
  $V_{H}$, and the asymptotic wind velocity exceeds the escape speed
  of the halo.  Mass loss rates are insensitive to the footpoint
  values of the magnetic field and angular momentum.  In addition to
  numerical parameter space exploration, we develop and compare to
  analytic scaling relations.  We show that winds have mass loss rates
  per unit area up to $\dot{\Sigma} \sim \Pi_{0}V_{H}^{-5/3}
  u_{0}^{2/3}$ where $\Pi_{0}$ is the footpoint cosmic ray
  pressure and $u_0$ is set by the upwelling of galactic fountains.
  The predicted wind mass-loss rate exceeds the star
  formation rate for $V_{H} \lesssim 200\kms$ and $u_{0} = 50\kms$, a typical
  fountain velocity.
\end{abstract}


\section{Introduction}\label{sec:intro}

The study of galactic winds seeks to understand the loss of mass from
galaxies. Mass loss through winds is believed to be responsible for
substantially reducing the observed baryon mass fraction in galaxies below
cosmic values, and for helping to quench ongoing star formation,
especially in low-mass galaxies
\citep[e.g.][]{Somerville2015,2017ARA&A..55...59N}.
Many studies have concluded that only up to $10-20\%$ of the cosmic
baryons can be found in stars and gas within galaxies
\citep[e.g.][]{Bell2003,2013MNRAS.428.3121M,
  2013ApJ...770...57B,2017MNRAS.470..651R}, and this fraction steeply drops
off for halos either above or below $\sim 10^{12} M_\odot$.
Except for the highest mass halos, the hot halo gas ($T\simgt 10^6\Kel$)
does not appear to make up for
the baryon deficit, but substantial warm ($T\sim 10^4\Kel$)
and warm-hot ($T\sim 10^5-10^6 \Kel$) gas
is present in circumgalactic regions for a range of halo masses and
redshifts, based on absorption-line surveys and other probes 
\citep[e.g.][]{Cen1999,Anderson2010,Chen2012,2012ARA&A..50..491P,2014ApJ...792....8W,2017ApJ...837..169P}.
As accretion timescales are shorter than the Hubble time, circumgalactic gas
that is accreted must subsequently be removed by galactic winds, and
these winds are also presumably responsible for enriching the
circumgalactic and intergalactic medium 
with metals \citep[e.g.][]{2017ARA&A..55..389T}.
                        
Direct evidence of winds from galaxies is given by high-velocity
emission and absorption lines that probe gas at a wide range of
temperatures \citep[see][for reviews]{Veilleux2005,Heckman2017}.
Most observations of galactic outflows have focused on starburst systems, and
indicate empirical scaling
relations that have yet to be fully explained. \cite{Martin2005} used
\ion{Na}{1} and \ion{K}{1} absorption lines in ultra-luminous infrared
galaxies to study cool gas outflows, finding that the outflow speed $v
\propto \dot M_*^{0.35}$, where $\dot M_*$ is the star formation
rate. With Cosmic Origin Spectrograph \textit{Hubble Space Telescope}
data from 48 nearby star-forming galaxies, \cite{Chisholm2015}
found that outflow velocities scale as
$v\propto \dot M_*^{0.08-0.22}$, $v\propto M_{\star}^{0.12-0.20}$, and
$v \propto v_{\mathrm{circ}}^{0.44-0.87}$, where $M_*$ is the total stellar content
and $v_{\rm circ}$ is the galaxy's circular velocity.
\cite{Chisholm2017} extended this
analysis to explore correlations between outflow rates and galaxy
properties for seven galaxies, finding a ratio of
mass outflow rate to star formation rate $\dot M_{\rm wind}/\dot M_* \equiv \beta$
(the ``mass loading factor'')
\begin{equation}
  \beta= 1.12 \pm 0.27 \left(\frac{v_{\mathrm{circ}}}{100 \mathrm{km/s}}\right)^{-1.56 \pm 0.25}.
\end{equation}
Work by \cite{Heckman2015} used ultraviolet absorption lines
in 39 galaxies to study warm ionized starburst-driven winds. \cite{Heckman2015}
found a slightly shallower power law for mass-loading
than \citet{Chisholm2017}, with best-fit
$\beta \propto v_{\mathrm{circ}}^{-0.98}$  for strong outflows, but their data are
also roughly consistent with power laws slopes between $-1$ and $-2$.
\cite{Heckman2016} found that outflow
velocities scale roughly as $v\propto \dot M_*^{0.3}$ and
$v\propto v_{\mathrm{circ}}^{1.16 \pm 0.36}$. The variations among recent 
reported observations suggests that empirical wind scaling relations are 
not yet definitive, and it is uncertain how these may extend from
starbursts to more
normal star-forming galaxies.

Proposed theoretical mechanisms for driving galactic winds have been
reviewed by \citet{Veilleux2005, Heckman2017}.  An important early
galactic wind model, motivated by the iconic starburst M82, considers
a hot, adiabatic radial flow that originates with specified mass and
energy input rates within the central region of a starburst nucleus
\citep{1985Natur.317...44C}.  The hot gas in models of this kind is
assumed to be created by extremely high velocity shocks arising from
stellar winds and supernovae.  The asymptotic velocity of the gas in
this model depends on the central gas temperature, which in turn
depends on the (adopted) ratio of energy to mass input rates.  If an
initially-hot wind of this kind has high enough energy loading to
reach high velocity, but also mass loading in the regime that allows
it to cool subsequent to acceleration, then radiative cooling by metal
lines could in principle produce a high velocity warm \second{or cold} outflow
\citep{1995ApJ...444..590W,2016ApJ...819...29B,Thompson2016}.
However, there is only a limited range of mass-loading $\beta_{\rm hot} \sim 1-2$
that allows a
wind to cool strongly {\it after} accelerating to high velocity
\citep{Thompson2016}, and it is not clear whether this range
\revise{of $\beta_{\rm hot}$} 
is compatible with the detailed interaction between blast waves from
multiple correlated supernovae and the surrounding interstellar medium (ISM).
\citet{2017ApJ...834...25K} show that except in extreme events, 
superbubbles are expected to cool
before breaking out of the surrounding ISM, and that the residual hot
gas at the time of breakout has $\beta_{\rm hot} \sim 0.1-1$.  Kim \&
Ostriker (2017, submitted) found in self-consistent simulations
(for Solar neighborhood conditions) with
star formation and supernova feedback that $\beta_{\rm hot} \sim 0.1$
above $z\sim 1 \kpc$, and the hot, high-velocity outflow remains adiabatic.

Another mechanism that has been proposed for driving a high-velocity
warm outflow is that a hot, high-velocity flow transfers momentum to
embedded warm \second{(or even cold)}, dense clouds.  A longstanding difficulty with this
cloud entrainment model, however, is that significant acceleration of
clouds is generally accompanied by cloud shredding and destruction on
short timescales \citep[e.g.][and references
  therein]{2015ApJ...805..158S,2017MNRAS.468.4801Z}.  Acceleration
of individual dense clouds by radiation pressure forces similarly
tends to destroy them \citep[e.g.][]{2014ApJ...780...51P,2017arXiv170802946Z}. 

Cosmic rays are believed to be accelerated in the shocks created by
supernovae, with $\sim 10\%$ of the injected energy going into cosmic rays,
and the local energy density of cosmic rays comparable to other energy
densities in the Milky Way's interstellar medium \citep[e.g.][]{2004MNRAS.353..550B,2015ARA&A..53..199G}.
GeV particles, which represent the largest contributor to the cosmic ray
energy density, are confined within the galaxy for only $\sim 10\Myr$, and in
flowing out of the galaxy they interact via the magnetic field with 
the ISM gas \citep[e.g.][]{2017PhPl...24e5402Z}.
Cosmic ray pressure gradients transfer momentum (and energy)
from the cosmic rays
to the gas, and may help to drive galactic winds.  In this paper,
we focus on analyzing the capability of cosmic ray-gas interactions 
to accelerate cool ($T\sim 10^4\Kel$) gas to high velocities such that
it is able to escape far into galactic halos.  

The first studies of a cosmic-ray-driven galactic wind were by
\cite{Ipavich1975}.  
He found that cosmic rays can drive galactic winds
with mass loss rates of $1-10 M_{\odot}/$yr, and that even zero
temperature gas can be accelerated.  A limitation of this exploratory
study was that the framework adopted was a spherical, Keplerian
potential in analogy to the solar wind.  As we shall show, the form of
the gravitational potential significantly affects the character of
winds, and in particular the potential associated with an extended
mass distribution in galaxies leads to constraints and types of wind
solutions that are quite different from those for a Keplerian
potential.

Further studies by \cite{Breit1991} incorporated a more realistic
galactic potential (Miyamoto-Nagai bulge-disk and dark matter halo),
and adopted the (arbitrary) assumption of vertical streamlines in which the
cross-sectional area varies as $A(z)/A_0=1+ (z/Z_0)^2$.  They focused on
non-radiative gas, and allowed for nonzero wave pressure.  From
their sampling of parameter space, they found that 
cosmic rays were necessary to drive a wind in many cases
(except for very high initial temperature), and in
particular, for typical conditions in the Milky Way galaxy.  Cases
with large initial (combined) energy density led to the highest mass-loss
rates, and higher initial density tended to reduce the mass-loss rate. 
\citet{Recchia2016} solved similar equations to \citet{Breit1991},
except that they assumed waves are fully damped, while allowing for
nonzero diffusivity that is self-consistently calculated based on
the wind solution.

\cite{Everett2008}, motivated by diffuse X-ray observations towards
the inner Galaxy, studied winds driven by a combination of cosmic ray
and thermal pressure.  They found that cases with cosmic ray pressure
comparable to the thermal gas pressure produced the best fit to the
observed Galactic diffuse soft X-ray emission. Their models indicate
that thermal pressure imparts momentum and energy to the flow early
on, and is more effective than cosmic ray pressure in mass-loading a
wind.  The terminal velocity and the evolution of the wind further
from the base is more sensitive to the cosmic ray pressure.
\revise{For fixed total (cosmic ray plus thermal) footpoint pressure,
  \cite{Everett2008} find that predominantly thermal-driven winds have
  higher mass-loss rates than predominantly cosmic-ray-driven winds.
  However, high thermal pressure is not guaranteed, and other work
  finds that the pressure of hot gas in the wind-launching region at
  $z \sim \kpc$ is insufficient to drive strong disk winds in typical
  star-forming galaxy environments (Kim \& Ostriker 2017,
  submitted).  }

In addition to idealized analytic models, three-dimensional
hydrodynamic and magnetohydrodynamic (MHD) simulations have recently been
performed to explore the role of cosmic ray pressure forces in driving
galactic outflows.  These have adopted varying assumptions concerning
the treatment of cosmic rays.  For example, \citet{Uhlig2012} do not
include diffusion or MHD, and assume that the cosmic ray fluid streams
at the sound speed along the direction of the cosmic ray pressure
gradient; \citet{Hanasz2013} and \citet{Simpson2016} neglect streaming
of the cosmic ray fluid relative to the gas but include advection at
the gas velocity and adopt fixed diffusion coefficients parallel and
perpendicular to the magnetic field; \citet{Booth2013} and
\citet{Salem2014a} and \citet{Salem2014b} neglect cosmic ray streaming and MHD,
adopting an isotropic diffusivity; \citet{Ruszkowski2017} compare
models in which cosmic rays stream along the cosmic ray pressure
gradient at a speed proportional to the Alfv\'en speed or diffuse
parallel to the magnetic field.  All of these simulation
studies have found that
cosmic ray pressure gradients can drive significant winds, with
mass-loss rates that can be comparable to star formation rates but are
dependent on the detailed prescription and parameters adopted.  A 
notable feature of simulations with a cosmic ray fluid is that galactic winds
include cool ($T \simlt 10^4\Kel$) gas.

In this paper, we extend steady state one-dimensional studies of cosmic-ray
driven winds to consider the case in which thermal pressure is negligible.
We are motivated by observations that suggest high-velocity cool winds
are ubiquitous (including even molecular gas), while at the same time 
simulations suggest that Type II supernovae interacting with
the ISM produce hot gas at a rate
$\dot M_{\rm hot}/\dot M_* = \beta_{\rm hot}\simlt 1$;
taken together, this argues that heavily mass-loaded winds ($\beta>1$)
must rely on acceleration of warm and cold (rather than hot) ISM phases
to speeds exceeding $\sim v_{\rm circ}$ that allow escape.  Thermal
pressure is included in our models by an isothermal equation of state
with $c_s=10\kms$, and it plays no role in wind acceleration.
\second{Although we do not explicitly follow the ionization level in the
gas, we implicitly assume that this is high enough for the gas to be
well-coupled to the magnetic field; for the low-density extraplanar warm
medium under consideration,
photoionization is believed to dominate
\citep[e.g.][]{2001RvMP...73.1031F,2009RvMP...81..969H}.
}  
We do not include cosmic ray diffusion or explicit wave pressure, assuming
that the cosmic ray fluid streams at the Alfv\'en speed relative to the
gas.  As in previous one-dimensional models, the streamline shape and
cross-sectional area are prescribed, but our choices for these follow from the
galactic potential rather than being arbitrary.  We integrate
the wind equation along streamlines to obtain the
gas velocity, density, magnetic field, and cosmic ray pressure, seeking
solutions that make smooth transitions through a sonic point.  To complement
our numerical solutions, we obtain analytic scaling relations
for the properties of winds.

In \autoref{sec:analysis} we
describe our assumptions and mathematical formulation
(\autoref{sec:hydro},\autoref{sec:stream}), derive a one-dimensional
steady wind equation
(\autoref{sec:wind}), discuss the critical point transition
and our integration method (\autoref{sec:sonic}), and
connect to a form of the Bernoulli equation (\autoref{sec:bernoullieq}).
Section \ref{sec:results} contains our results. 
We specify the details of our galactic models
and input parameterization (\autoref{sec:models}),
give examples of wind solutions for dwarf and Milky Way galaxies
(\autoref{sec:sample}), and present
results from our full parameter exploration of solutions
to the wind equation (\autoref{sec:param}).  In 
\autoref{sec:scaling} we derive analytic scaling relations 
for wind properties, and compare to our numerical integrations.  
Section \ref{sec:angmag} explores the effects of varying angular momentum
and magnetic field strength on wind solutions.
The key output of our study is a theoretical prediction for the
mass-loss rates and mass-loading factors of cosmic-ray driven disk winds,
which we discuss in \autoref{sec:mdot}.
Finally, \autoref{sec:sum} summarizes and discussed our main conclusions.  
\revise{In \autoref{sec:ind} we provide estimates for the effect of ion-neutral
 collision-induced wave  damping on the cosmic ray streaming speed, and in}
\autoref{sec:ceff2} we provide additional details related to
the behavior of the effective sound speed.

\section{Analysis}\label{sec:analysis}

\subsection{Hydrodynamic Equations}\label{sec:hydro}

We begin with the equations governing the combined gas and cosmic ray
fluid flow \citep[e.g.][]{Breit1991}. The fluid variables are gas density
$\rho$, 
gas velocity $\vect{v}$, gas pressure $P$, gas internal energy
density $\mathcal{E}$, magnetic field $\vect{B}$, 
cosmic ray pressure $\Pi$, and cosmic ray energy
density $\mathcal{E}_{\rm cr}$.  The
Alfv\'en velocity is given by  $\vect{v}_{A} = \vect{B}/(4\pi \rho)^{1/2}$,
and the total gravitational potential, including
both stars and dark matter, is $\Phi$.
The
collective flow of cosmic rays along the magnetic field is limited by
the streaming instability, in which a mean cosmic ray velocity
\revise{(relative to the gas)} 
exceeding the Alfv\'en speed leads to resonant excitation of
Alfv\'en waves that then pitch-angle scatter the cosmic rays
\citep{1969ApJ...156..445K}.  We assume that wave damping keeps the
amplitude of excited waves low, and also mediates the transfer of
momentum from the cosmic ray fluid to the gas
\citep[e.g.][]{2005ppa..book.....K,2017PhPl...24e5402Z}.  Although
very efficient wave damping can lead to faster streaming
\second{\citep{2011ApJ...739...60E,Wiener2013,Wiener2017}}, we shall assume that the mean velocity of the cosmic
ray distribution in the rest frame of the gas is equal to $\vect{v}_A$.
\footnote{
  The drift speed $v_{D}$ of the cosmic ray fluid relative to the gas
  depends on the damping mechanism for the Alfv\'en waves responsible
  for pitch-angle scattering.  
  \cite{Wiener2017} Equations (6) and (7)
  respectively provide estimates for the streaming speed under assumptions
  of nonlinear Landau damping and turbulent damping.  
  In both cases,  $v_{\rm D} \sim v_{A}$
  for the parameter regime we consider, consistent with our assumption.
  Ion-neutral collisions can further damp waves, but for coronal regions
  where galactic winds originate the neutral density is low and wave
  damping is weak (see \autoref{sec:ind}).
}

We adopt a cylindrical coordinate system with unit vectors
$\hat{R}$, $\hat{z}$, and $\hat{\phi}$, with $z=0$ in the midplane of the
galactic disk.  We take $\Omega$ as the local mean rotational velocity of ISM
gas in the disk where the wind originates.  
The inertial-frame velocity $\vect{v}$ is related to the velocity $\vect{u}$ 
in a frame rotating with angular velocity $\Omega \hat z$ by
\begin{equation}
\vect{v} = \vect{u} + \Omega R\hat{\phi}.
\end{equation}
Mass conservation is expressed by  
\begin{equation}\label{eq:continuity}
\partial_{t}\rho + \nabla\cdot(\rho\vect{v}) = 0.
\end{equation}
The momentum equation for the gas in the inertial frame is 
\begin{equation}
\partial_{t}\vect{v} + \vect{v}\cdot\nabla\vect{v} + \frac{\nabla{}P}{\rho} + \frac{\nabla\Pi}{\rho} = -\nabla\Phi + \frac{(\nabla\times\vect{B})\times\vect{B}}{4\pi\rho},
\end{equation}
 which becomes  
\begin{equation}\label{eq:momentum}
\begin{split}
  & \partial_{t}\vect{u} + \vect{u}\cdot\nabla\vect{u} + 2\vect{\Omega}\times\vect{u} + \frac{\nabla(P+\Pi)}{\rho}
  = -\nabla\left(\Phi - \frac{\Omega^{2}R^{2}}{2}\right) + \frac{(\nabla\times\vect{B})\times\vect{B}}{4\pi\rho}
\end{split}
\end{equation}
in the rotating frame.

Assuming that the cosmic ray fluid streams along the magnetic field at velocity
$\vect{v}+\vect{v}_A$, and that cosmic ray
diffusion and radiative and collisional energy losses may
be neglected, the energy equation for the cosmic ray fluid is 
\begin{equation}
\begin{split}
  & \partial_{t} \mathcal{E}_{\rm cr} 
  + \nabla \cdot [(\vect{v} + \vect{v_{A}})(\mathcal{E}_{\rm cr} + \Pi) ]
  = (\vect{v} + \vect{v_{A}}) \cdot \nabla\Pi.
\end{split}
\end{equation}
Note that $\vect{v} \cdot \nabla\Pi$ represents the
work done by the cosmic ray fluid in accelerating the gas,
and $\vect{v_{A}}\cdot \nabla\Pi$
represents energy losses due to generation of Alf\'ven waves.

The \revise{general form for the}
internal energy equation for the gas is given by 
\begin{equation}\label{eq:internalee}
\begin{split}
  & \partial_{t} \mathcal{E} 
  + \nabla \cdot[\vect{v}(\mathcal{E} + P) ]
  = \vect{v}\cdot \nabla P - \vect{v_{A}} \cdot \nabla\Pi - \rho {\cal L},
\end{split}
\end{equation}
where $\rho\mathcal{L}$ is the net radiative loss per volume per time.
The term $\vect{v}\cdot \nabla P$ represents
work done in accelerating the flow, while the term
$-\vect{v_{A}} \cdot \nabla\Pi$ represents heat energy gained by
wave damping.

For $\mathcal{E}_{\rm cr} = \Pi/(\gamma_{\rm cr}-1)$, ${\cal E} = P/(\gamma -1)$, 
and an axisymmetric flow, the cosmic ray and gas \revise{thermal} energy equations become 
\begin{equation}\label{eq:crenergy}
  \partial_{t} {\Pi} + (\vect{u} + \vect{v_{A}}) \cdot \nabla{\Pi} +
  \gamma_{\rm cr} \Pi \nabla\cdot(\vect{u} + \vect{v_{A}}) = 0
\end{equation}
and 
\begin{equation}\label{eq:gasenergy}
\begin{split}
  \partial_{t} P + \vect{u}\cdot\nabla{}P
  + \gamma P \nabla\cdot \vect{u} =
  -(\vect{v_{A}} \cdot \nabla\Pi + \rho\mathcal{L})(\gamma -1).
\end{split}
\end{equation}

\revise{Previous steady state wind solutions adopt Equations \ref{eq:continuity}, \ref{eq:momentum}, \ref{eq:crenergy}, and \ref{eq:gasenergy} with $\partial_{t} = 0$, usually also taking $\mathcal{L} = 0$.}

We are interested in winds consisting of warm gas
that is maintained at $T\sim 10^4\Kel$ by radiative + shock
heating and radiative cooling.  \revise{Rather than implementing
  gas heating and cooling terms, for simplicitly we instead} adopt an isothermal
equation of state \revise{with $P = \rho c_{s}^{2}$ along streamlines for $c_{s}$ the constant sound speed. This is equivalent to $\gamma=1$  in
  \autoref{eq:gasenergy} because 
$\nabla \cdot \vect{u} = - \vect{u}\cdot \nabla \ln\rho$ from the
  continuity equation.}
  \footnote{
    \revise{We note that when $\gamma \ne 1$, for ${\cal L}=0$ the cosmic
      ray energy equation, gas thermal
      energy equation, and momentum equation (dotted
      with $\rho \mathbf{v}$) can be combined to
      obtain an equation expressing total energy conservation in the flow,
      $\nabla \cdot\left[ \frac{1}{2}v^2 \rho \mathbf{v} +
        \frac{\gamma}{\gamma -1}P\mathbf{v} + \Phi \rho\mathbf{v}+
\frac{\gamma_{\rm cr}}{\gamma_{\rm cr} -1}\Pi (\mathbf{v}+ \mathbf{v}_A)   
\right]=0$,
      which is related to the Bernoulli equation.
      While this expression does not apply when $\gamma =1$, a
      different  Bernoulli-like equation can be obtained in that case
    (see \autoref{sec:bernoullieq}).}  }
  
We assume that Lorentz forces $(\nabla\times\vect{B})\times\vect{B}$ are
negligible, so that in axisymmetry the $\hat \phi$
component of Equation (\ref{eq:momentum}) implies angular momentum is
conserved along each streamline,
\begin{equation}\label{eq:angmomentum}
v_\phi R = (u_\phi + \Omega R)R = J =const.
\end{equation}
With $\vect{u}_p=u_R \hat R + u_z \hat z$ the poloidal velocity, the
poloidal components of Equation (\ref{eq:momentum}) becomes
\begin{equation}\label{eq:polmomentum}
  \vect{u}_p\cdot \nabla \vect{u}_p + \frac{\nabla (P+\Pi)}{\rho} =
  -\nabla \Phi + \frac{v_\phi^2}{R}=-\nabla \Psi,
\end{equation}  
where the effective potential
\begin{equation}
\Psi \equiv \Phi + \frac{J^2}{2R^2}
\end{equation}
incorporates centrifugal-force effects.  

\subsection{Flow Streamlines and Conserved Quantities}\label{sec:stream}

A major assumption in this work is that the poloidal components of
the fluid and Alfv\'en velocities and the gradients of the pressures 
are all aligned with the gradient of the effective
gravitational potential $\nabla\Psi$.  
We assume that all these vectors lie along
$\hat{s}$, the streamline direction. For streamlines in the poloidal
(R-z) plane, the tangent direction is 
\begin{equation}\label{eq:tangent}
\hat{s} = \frac{(dR/dz)\hat{R} + \hat{z}}{[(dR/dz)^{2} + 1]^{1/2}}.
\end{equation}
The normal to the streamline in the poloidal plane is given by
\begin{equation}
\hat{t} = \frac{\hat{R} - (dR/dz)\hat{z}}{[(dR/dz)^{2} + 1]^{1/2}}.
\end{equation}
Since $\hat{s}$ lies along the gradient of $\Psi$, 
\begin{equation}
\hat{t} \cdot \nabla\Psi = 0, 
\end{equation}
and the streamline can be found from the potential by solving 
\begin{equation}\label{eq:drdz}
\frac{dR}{dz} = \frac{\partial\Psi/\partial{}R}{\partial\Psi/\partial{}z}. 
\end{equation}
The distance $s$ along the streamline is obtained from
\begin{equation}\label{eq:dsdz}
\frac{ds}{dz} = \left[1 + \left(\frac{dR}{dz}  \right)^2  \right]^{1/2}.
\end{equation}  
The area $A$ of a given fluid element (or the axisymmetric area $A$
between two poloidal streamlines) varies with $s$ as 
\begin{equation}\label{eq:dslna}
d_{s} \ln{A} = \nabla\cdot\hat{s}, 
\end{equation}
where the right-hand side is obtained from applying the divergence to
Equation (\ref{eq:tangent}) with Equation (\ref{eq:drdz}).  As an example,
radial streamlines have $\hat s =\hat r$, and
$d_r \ln A = \nabla \cdot \hat r =2/r$ so that $A\propto r^2$.  
If $z$ is taken
as the independent variable, we instead have
\begin{equation}\label{eq:dlnAdz}
  d_z\ln{A}= \frac{ds}{dz}d_s\ln{A},
\end{equation}
  and use
Equation (\ref{eq:dsdz}).

\begin{figure}
\includegraphics[width=\columnwidth]{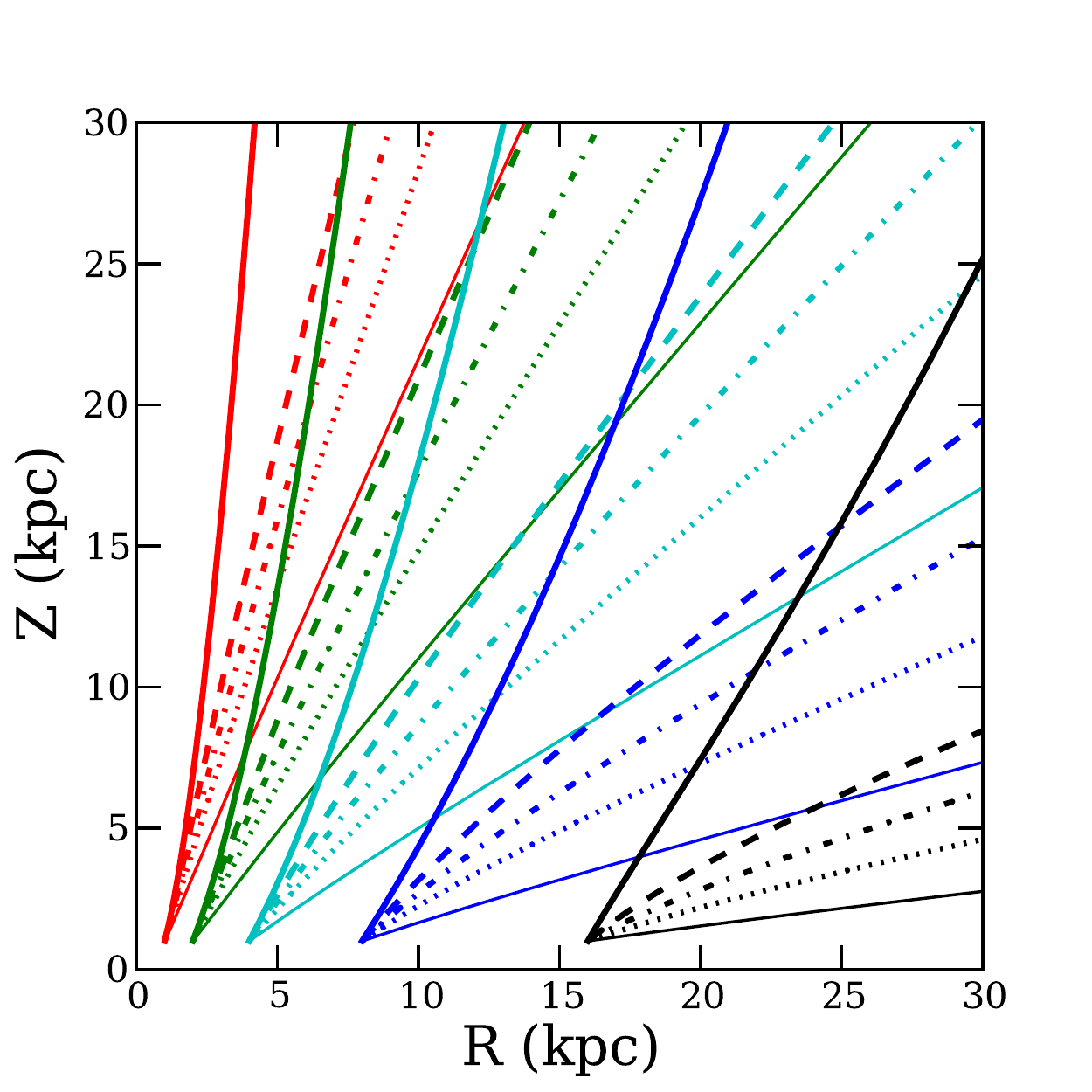}
\caption{A set of example streamlines starting at $z = 1\kpc$ 
  and $R_0 = 2^{N}$ kiloparsecs. The different line styles correspond to
  different values of $J$ within a representative Milky Way
  potential with $V_H = 250 \kms$ and $R_{\rm vir} = 250 \kpc$. 
  A larger value of $J$ provides an outward acceleration
  which counters some of the inward gravitational acceleration, which
  causes the streamline to be more vertical. See text for details. }
\label{fig:streamline}
\end{figure}

Figure \ref{fig:streamline} shows examples of streamlines emerging
from the disk for a Milky Way potential $\Phi$, for a range of values
of $J$ in $\Psi$.  For each footpoint the five values of $J$
correspond to 0, 0.8, 0.9, 0.95, and 1.0 times the respective \revise{maximum} value
on each footpoint.  
These \revise{maximum} values correspond to the angular momentum of a
circular orbit at radii of 0.59, 1.39, 2.99, 6.74, and 15.14 kpc, 
respectively, \revise{for a halo with virial radius 250 kpc. These values scale with the virial radius.} See \autoref{sec:models} for details regarding the
potential, and \autoref{sec:angmag} for a discussion of $J$ and
definition of the \revise{maximum} value.

Henceforth, we use $u$ to denote the magnitude of the poloidal gas
velocity, with
\begin{equation}\label{eq:polvel}
  \vect{u}_p=u \hat s,
\end{equation}
and similarly
\begin{equation}\label{eq:polvA}
  \vect{v}_{A,p}=v_A \hat s.
\end{equation}

From mass conservation (\autoref{eq:continuity}),
$ \nabla \cdot (\rho u \hat s)=0$, which implies
$d_{s}\ln\rho = - (d_{s}\ln{u} + d_{s} \ln{A})$.
Thus,
\begin{equation}\label{eq:rho}
  \rho=  \rho_0\left(\frac{uA}{u_{0}A_{0}}\right)^{-1} =
  \rho_c \left(\frac{uA}{u_{c}A_{c}}\right)^{-1}
\end{equation}
where the ``$0$'' subscript denotes values at the streamline footpoint,
and the ``$c$'' subscript denotes values at the streamline critical point (see
\autoref{sec:sonic} for a discussion of critical points).

Similarly, the cosmic ray energy equation (\autoref{eq:crenergy}) becomes 
\begin{equation}\label{eq:pi}
  \Pi= \Pi_0 \left(\frac{u+v_{A}}{u_{0}+v_{A,0}}\frac{A}{A_{0}}\right)^{-\gamma_{\rm cr}}
= \Pi_c \left(\frac{u+v_{A}}{u_{c}+v_{A,c}}\frac{A}{A_{c}}\right)^{-\gamma_{\rm cr}}; 
\end{equation}
\revise{for $n_{\rm cr}$ the cosmic ray number density, this is consistent
with conservation of the flow of cosmic ray particles,
$(u+ v_A)A n_{\rm cr}=const.$, together with the relation
$\Pi \propto n_{\rm cr}^{\gamma_{\rm cr}}$.}

Since $\nabla\cdot\vect{B} = 0$ and $\vect{B}_p = B\hat{s}$, $d_{s}\ln{B} = -d_{s}\ln{A}$ so $B \propto A^{-1}$ and the Alfv\'en speed 
$v_{A} = B/\sqrt{4\pi\rho}$ evolves as 
\begin{equation}\label{eq:va}
  v_A= v_{A,0}\left(\frac{u/A}{u_{0}/A_0}\right)^{1/2}
=v_{A,c}\left(\frac{u/A}{u_{c}/A_c}\right)^{1/2}.
\end{equation}
This expresses the combined conservation of magnetic flux and mass flux.

Note that the ratio of the Alfv\'en speed to the wind speed evolves as 
\begin{equation}\label{eq:vau}
  \begin{split}
\frac{v_{A}}{u} &= \frac{v_{A,0}}{u_{0}} \left(\frac{uA}{u_0A_0}\right)^{-1/2}
               = \frac{v_{A,0}}{u_{0}} \left(\frac{\rho}{\rho_{0}}\right)^{1/2}\\
               &= \frac{v_{A,c}}{u_{c}} \left(\frac{uA}{u_cA_c}\right)^{-1/2}
               = \frac{v_{A,c}}{u_{c}} \left(\frac{\rho}{\rho_{c}}\right)^{1/2}.
\end{split}
  \end{equation}
For an accelerating wind whose streamlines are opening, both $u$ and
$A$ monotonically increase with $s$ while $\rho$ decreases, so $v_{A}/u$ must
decrease with increasing $s$.

\subsection{One-dimensional Steady Wind Equation}\label{sec:wind}

Applying the assumptions described in \autoref{sec:hydro} and \autoref{sec:stream} to \autoref{eq:polmomentum}, the poloidal momentum equation becomes 

\begin{equation}
  u d_{s} u + c_{s}^{2} d_{s}\ln{\rho} -
  \gamma_{\rm cr}\frac{\Pi}{\rho}d_{s} \ln{\left[(u+v_{A})A\right]} = -d_{s}\Psi.
\end{equation}

After some manipulation, we find

\begin{equation}
\begin{split}
  & \left(u^{2} - c_{s}^{2} - \gamma_{\rm cr} \frac{\Pi}{\rho}\frac{u+v_{A}/2}{u+v_{A}}\right)d_{s}u
  = u\left(-d_{s}\Psi + c_{s}^{2}d_{s}\ln{A} + \gamma_{\rm cr}\frac{\Pi}{\rho}\frac{u+v_{A}/2}{u+v_{A}}d_{s}\ln{A}\right).
\end{split}
\end{equation}

We define an effective sound speed $C_{\rm eff}$, including effects of
both gas and cosmic ray pressure, by the expression  
\begin{equation}\label{eq:ceff2}
C_{\rm eff}^{2} \equiv c_{s}^{2} + \gamma_{\rm cr}\frac{\Pi}{\rho}\frac{u+v_{A}/2}{u+v_{A}}
\end{equation}
(cf. Equations 30 and 31 of \citealt{Ipavich1975}).

Starting from \autoref{eq:pi} and using \autoref{eq:rho} and \autoref{eq:vau},
one can show that
\begin{equation}\label{eq:dpidrho}
\frac{d\Pi}{d\rho}=\gamma_{\rm cr} \frac{\Pi}{\rho}\frac{u+ v_A/2}{u+v_A}.
\end{equation}  
Thus, $C_{\rm eff}^2$ in \autoref{eq:ceff2} can also be written as
\begin{equation}\label{eq:ceff2alt}
  C_{\rm eff}^2 = \frac{dP}{d\rho} + \frac{d\Pi}{d\rho}.
\end{equation}

We define a gravitational velocity $V_{g}$ by the expression 
\begin{equation}\label{eq:gvel}
V_{g}^{2} \equiv \frac{d_{s}\Psi}{d_{s}\ln{A}}.
\end{equation}
We note that if streamlines are radial and the centrifugal term in
$\Psi$ is negligible, $V_g^2=rd_r\Phi/2 = v_c^2(r)/2$ for $v_c(r)$ the circular
velocity at distance $r$.  Thus, if the circular velocity is a nearly
constant value characterized by the galaxy's dark matter
halo, $V_g \sim V_H/\sqrt{2}$.  

With the above definitions, the ordinary differential equation that
describes the steady-state wind is given by 
\begin{equation}\label{eq:winds}
d_{s} u = u \frac{\left(V_{g}^{2}-C_{\rm eff}^{2}\right)}{\left(C_{\rm eff}^{2} - u^{2}\right)} d_{s}\ln{A}
\end{equation}
Written in this way, the wind equation (\autoref{eq:winds}) 
has the same form as that of
a classical Parker wind in a Keplerian potential,
taking $s\rightarrow r$,
$C_{\rm eff}^2 \rightarrow dP/d\rho\equiv c_s^2$, $V_g^2 \rightarrow (1/2)G M/r$,
and $d_s \ln A \rightarrow 2/r$.
In the case of general rather than radial streamlines, it is
convenient to use $z$ rather than $s$ as the independent variable, in
which case the wind equation may be written
\begin{equation}\label{eq:wind}
d_{z} u = u \frac{\left(V_{g}^{2}-C_{\rm eff}^{2}\right)}{\left(C_{\rm eff}^{2} - u^{2}\right)} d_{z}\ln{A}. 
\end{equation}

We note that
the density $\rho$ (or gas pressure $\rho c_s^2$)
appears in the wind equation only in ratios with the magnetic
pressure (in $v_A$) and the cosmic ray pressure.
For integration of the wind equation, we therefore only require the combination 
\begin{equation}\label{eq:pirho}
  \begin{split}
\frac{\Pi}{\rho} &= \frac{\Pi_0}{\rho_0} \left(\frac{u+v_A}{u_0 + v_{A,0}}\frac{A}{A_0}  \right)^{-\gamma_{\rm cr}} \left(\frac{u A}{u_0 A_0}\right) \\
                 &=\frac{\Pi_c}{\rho_c} \left(\frac{u+v_A}{u_c + v_{A,c}} \frac{A}{A_c} \right)^{-\gamma_{\rm cr}} \left(\frac{u A}{u_c A_c}\right)
\end{split}
\end{equation}
rather than \autoref{eq:rho} and \autoref{eq:pi} separately.

To obtain wind solutions, we
evolve $u$, $A$, and the streamline using \autoref{eq:wind},
\autoref{eq:dlnAdz}, and \autoref{eq:drdz} as a set of three coupled
ordinary differential equations.
For any point on the streamline where
we have $u$ and $A$, we find $v_A$ and $\Pi/\rho$ in terms of
$u$ and $A$ via \autoref{eq:va} and \autoref{eq:pirho}, respectively.

\subsection{Critical Point and Integration Method}\label{sec:sonic}

A physically realistic wind begins close to the galactic disk from a
velocity $u$ that is low compared to the effective sound speed $C_{\rm eff}$
and the gravitational speed $V_{g}$.  From \autoref{eq:winds},
for an accelerating wind with
$d_{s}u > 0$, it must be true that $V_{g} > C_{\rm eff}$ for $u < C_{\rm eff}$,
and $V_{g} < C_{\rm eff}$ for $u > C_{\rm eff}$.  
If the fluid is to achieve speeds that will allow it to escape into
the galaxy's halo, $u$ must exceed both $C_{\rm eff}$ and  $V_{g}$.
Since $d_{s}\ln{A}$ is set by the shape of the potential
$\Psi$, it is in general non-zero. Thus, for the flow to avoid
singularities (i.e. $d_{s}u$ is never infinite), at 
the critical point where $u=C_{\rm eff}$ it must also be true that
$V_{g} = C_{\rm eff}$.

\begin{figure*}
\includegraphics[width=\textwidth]{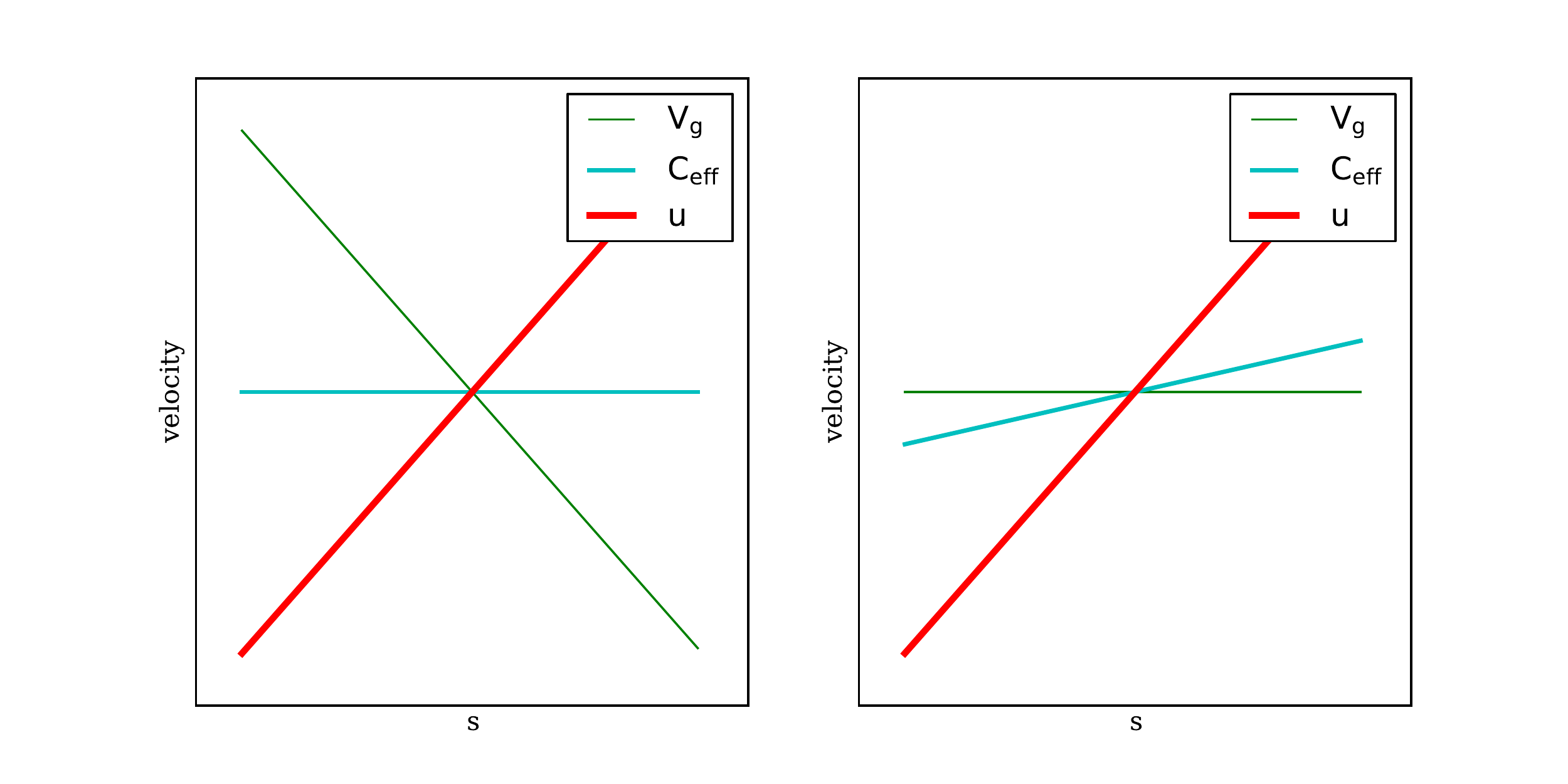}
\caption{A schematic comparing the behavior near the critical 
  point of an isothermal Parker wind in a Keplerian potential (left)
  to that of a wind driven by cosmic ray pressure
  in a galactic potential (right). \second{Loci} 
  depicting the gravitational velocity $V_{g}$, effective sound speed
  $C_{\rm eff}$, and wind velocity $u$ as a function of distance along
  the streamline $s$ are shown. Note that all three curves intersect at
  the  critical point (or sonic point).
  Also note that in each case,
  $u < C_{\rm eff} < V_{g}$ at low $s$, and $V_{g} < C_{\rm eff} < u$ at
  high $s$, consistent with the wind equation 
(\autoref{eq:winds}) for an accelerating
  flow.
  For the schematic Parker wind depicted, $C_{\rm eff}$ is constant
  (isothermal) and $V_{g}$ is
  decreasing (Keplerian). For the schematic cosmic ray driven wind, $V_{g}$ is
  nearly constant (galactic potentials have close to flat rotation curves),
  which necessitates an increasing $C_{\rm eff}$ to enable a critical transition.
  More generally, if
  $P_{\rm e} \propto \rho^{\gamma_e}$, $C_{\rm eff}^2 \equiv dP_{\rm e}/d\rho \propto
  \rho^{\gamma_e -1}$, so a critical transition is only possible in $V_g\sim const$
  galactic potentials if $\gamma_e <1$, whereas critical transitions
  are possible in Keplerian potentials ($V_g \propto r^{-1/2}$) for
  $\gamma_e \ge 1$.}
\label{fig:cartoon}
\end{figure*}

From \autoref{eq:ceff2}, one can show (see \autoref{sec:ceff2}) that
\begin{equation}\label{eq:dsceff2}
  \mathrm{sgn}(d_s C_{\rm eff}^2) = \mathrm{sgn}(d_s \rho)
  \mathrm{sgn}\left[
    \gamma_{\rm cr}\left(1 + \frac{1}{2}\frac{v_A}{u}\right)^2 -
    1 - \frac{7}{4} \frac{v_A}{u} - \frac{1}{2}\frac{v_A^2}{u^2}
      \right].
\end{equation}
For $\gamma_{\rm cr}=4/3$, $C_{\rm eff}$ will increase outward (as $\rho$ decreases)
whenever $v_A/u > 0.64$. From \autoref{eq:vau}, $v_A/u$ is strictly
decreasing with $s$ if $\rho$ is decreasing, so provided that
$v_A/u>0.64$ at the
critical point, $C_{\rm eff}$ will secularly increase from the footpoint up
to the critical point.  A schematic showing
$u$, $C_{\rm eff}$, and $V_{g}$ relative to one
another as a function of streamline distance, including a critical
transition, is shown in \autoref{fig:cartoon}.\footnote{
We discuss in \autoref{sec:scaling} the key differences between a Parker-type
  stellar winds driven
  by thermal pressure in a point mass potential, versus galactic winds
  driven by cosmic ray pressure in an extended dark matter halo potential.}

For a given galactic potential and streamline shape, the location of
the critical transition $R=R_{c}$, $z=Z_{c}$ fully specifies the value
of $V_{g,c}$.  Thus, a given location for the critical point also
specifies the fluid velocity $u_c$ and value of $C_{\rm eff,c}$ at
that point.

We obtain wind solutions to our set of ODEs with the following procedure:
Given some desired
footpoint ($R_{0},Z_{0}$) in the effective potential, we pre-compute
the streamline which passes through that footpoint by integrating
\autoref{eq:drdz} outward. Then, one may choose
some point ($R_{c},Z_{c}$) along that streamline to be the critical point;
this also specifies the values of $V_{g,c}=u_c=C_{\rm eff,c}$
and $A_c$ based on the potential and
streamline shape at the critical point.
Given ($R_{c},Z_{c}$), one may select a value of the Alfv\'en speed at the
critical point, $v_{A,c}$.  Then, applying \autoref{eq:ceff2} at the
critical point yields $\Pi_c/\rho_c$ in terms of $V_{g,c}$, $v_{A,c}$,
and $c_s$.
With all the fluid variables known at the critical point, the coupled ODEs
may be integrated back to the footpoint  
($R_{0},Z_{0}$) according to the procedure described at the end of
\autoref{sec:wind}. When the streamline
footpoint is reached, the starting ``ISM conditions''
$u_0$, $v_{A,0}$, and $\Pi_0/\rho_0$ that are consistent with the selected
critical point are read off of the solution.

For each footpoint, a variety of solutions can be
attained by (1) varying the critical point location ($R_{c},Z_{c}$) along
the streamline,
and (2) varying the Alfv\'en velocity at the critical point, $v_{A,c}$. 
In total, this implies two degrees of freedom for each footpoint and
streamline shape.  Equivalently, two degrees of freedom also represents
choosing the footpoint values of 
$\Pi_{0}/\rho_{0}$ and $v_{A,0}^2=B^{2}_{0}/(4 \pi \rho_{0})$,
with $u_{0}$ the unique value for which a solution is able to pass
through a critical point.  
Thus, we can explore
a range of ISM properties given a footpoint, and can use 2-D
root-finding to locate wind solutions \revise{(including the value $u_0$)}
of particular points in
$\Pi_{0}/\rho_{0}$ and $B^{2}_{0}/(4 \pi \rho_{0})$ space, while using the
critical point location and $v_{A,c}$ as inputs.
\revise{More generally, any two of the
  three footpoint velocities $(\Pi_0/\rho_0)^{1/2}$, $v_{A,0}$, $u_0$
  can be chosen to parameterize the space of possible solutions,
  with the third velocity constrained by the requirement that the flow
  makes a critical transition.
}

To initiate integration near the critical point, we apply 
L'H\^{o}pital's rule to the right-hand side of \autoref{eq:winds}:
\begin{equation}\label{eq:lhopital}
  \begin{split}
d_{s}u = \lim_{s\to s_{c}} u \frac{\left(V_{g}^{2}-C_{\rm eff}^{2}\right)}{\left(C_{\rm eff}^{2} - u^{2}\right)} d_{s}\ln{A}
  &= u_c d_{s}\ln{A} \frac{\frac{d}{ds}\left(V_{g}^{2}-C_{\rm eff}^{2}\right)}{\frac{d}{ds}\left(C_{\rm eff}^{2} - u^{2}\right)}
  \\
&= u_c d_{s}\ln{A}\frac{\left[d_{s}V_{g}-(\partial_{A}C_{\rm eff}d_{s}A + \partial_{u}C_{\rm eff}d_{s}u)\right]}{\left[\partial_{A}C_{\rm eff}d_{s}A + \partial_{u}C_{\rm eff}d_{s}u - d_{s}u\right]},
  \end{split}
\end{equation}
  where we use $u_c = V_{g,c} = C_{\rm eff,c}$ at the critical point and the partial derivatives with respect to $u$ assume holding $A$ constant and vice versa. Note that $C_{\rm eff}$ can be written as a function of $u$ and $A$. 
  
This yields a quadratic which must be solved for $d_{s}u$, after
computing $\partial_{A}C_{\rm eff}$, $\partial_{u}C_{\rm eff}$, $d_{s}A$, and
$d_{s}V_{g}$  (see \autoref{sec:ceff2}).
The two possible solutions are a decelerating
wind and accelerating wind, and the accelerating solution is
taken. Alternatively, using the properties of the solution topology,
different values of $f(u,s) = d_{s}u $ can be tested. Each value of
$d_{s}u$ will result in some $u' = u_{c} - d_{s}u \Delta{}s$ for a new
point $s' = s_{c} - \Delta{}s$. Then, taking this value of $u'$ and
position on the streamline $s' = s_{c} - \Delta{}s$, the derivative
$f(u',s')$ can be calculated. The true $f(u,s) = d_{s}u$ will be a
fixed point such that $f(u,s) = f(u',s') = f(u -
f(u,s)\Delta{}s,s-\Delta{}s)$ and can be numerically found.  This only
holds true for the true wind solution passing through the critical
transition, and does not hold true for the breeze solutions, due to
the solution topology of wind flows. Any error in this technique is
comparable to a shooting technique error, as even an order unity error
in $d_{s}u$ leads to a point within $\Delta{}s d_{s}u$ of the critical
point. That is, we begin near the sonic point in ($s$,$u$) space, as
long as the initial step $\Delta{}s$ is chosen to be small, which
avoids the sensitive nature of $d_{s}u$ near the sonic point and gives
us an accurate wind. Integration can proceed directly from there.

\subsection{Bernoulli Equation}\label{sec:bernoullieq}

From \autoref{eq:rho}, \autoref{eq:gvel}, and \autoref{eq:winds} 
it is straightforward to show that
\begin{equation}\label{eq:bernoulli}
d_{s} \left(\frac{1}{2}u^{2} + \Psi\right) = -C_{\rm eff}^{2} d_{s}\ln{\rho}.
\end{equation}

\revise{With \autoref{eq:ceff2alt}}, \autoref{eq:bernoulli}  then becomes
\begin{equation}
  d_{s} \left(\frac{1}{2}u^{2} + \Psi\right) =
  -\left(\frac{d_sP}{\rho} + \frac{d_s\Pi}{\rho}  \right).
\end{equation}
For an isothermal equation of state for the gas, $dh_g=dP/\rho$ for gas enthalpy
$h_g=c_s^2 \ln\rho$.  We can formally define cosmic ray enthalpy $h_{\rm cr}$ via
$dh_{\rm cr}=d\Pi/\rho$.  With this definition we have
\begin{equation}
d_s  \left(\frac{1}{2}u^{2} + \Psi + h_g + h_{\rm cr}\right) \equiv d_s{\cal B} =0
\end{equation}
for Bernoulli parameter $\cal B$.
In general, \autoref{eq:dpidrho} does
not yield a simple analytic form for $h_{\rm cr}$.  However, in the limit of
either $u\gg v_A$ or $u\ll v_A$ we have $\Pi \propto \rho^{\gamma_{\rm cr}}$ or
$\Pi \propto \rho^{\gamma_{\rm cr}/2}$, respectively, such that 
\begin{equation}
  h_{\rm cr} \rightarrow
  \begin{cases}
    \frac{\gamma_{\rm cr}}{\gamma_{\rm cr} -1}\frac{\Pi}{\rho}& \ \  for \ \ u\gg v_A \\
     \frac{\gamma_{\rm cr}}{\gamma_{\rm cr} -2}\frac{\Pi}{\rho}& \ \ for \ \ u\ll v_A
  \end{cases}
\end{equation}
in the two limiting cases.

With $\gamma_{\rm cr} =4/3 $, $h_{\rm cr} \rightarrow 4(\rho/\rho_0)^{1/3} \Pi_0/\rho_0$
and $C_{\rm eff}^2 \rightarrow (4/3)(\rho/\rho_0)^{1/3} \Pi_0/\rho_0$
for $u\gg v_A$, while $h_{\rm cr} \rightarrow -2(\rho/\rho_0)^{-1/3} \Pi_0/\rho_0$
and $C_{\rm eff}^2 \rightarrow (2/3)(\rho/\rho_0)^{-1/3} \Pi_0/\rho_0$
for $u\ll v_A$.  The case $u\gg v_A$ has the same characteristic
behavior as gas enthalpy, in that $h_{\rm cr}$ is positive and both
$h_{\rm cr}$ and $C_{\rm eff}$ decrease
in magnitude as $\rho$ decreases.  The limit $u\ll v_A$, which is more relevant
for understanding wind solutions
inside the critical point, has instead very different behavior:
$h_{\rm cr}$ is negative, and both
$h_{\rm cr}$ and $C_{\rm eff}$
increase in magnitude as $\rho$ decreases.  It is
this  behavior for $h_{\rm cr}$ and $C_{\rm eff}$
that allows $u$ to increase and smoothly pass
through a critical point where $u=V_g=C_{\rm eff}$ even when $V_g$ is nearly
flat in $s$ (see \autoref{fig:cartoon}).

\section{Results}\label{sec:results}

\subsection{Model Specification}\label{sec:models}

Our goal is to explore the dependence of possible wind properties, and
especially mass-loss rates, on the galactic environment.
Winds will be affected by both the properties of the ISM
in which the wind originates, and the galactic potential in which it is
accelerated.

To represent a range of galactic potentials, we adopt the general form of 
\cite{Bovy2015} for the Milky Way potential.  This includes 
a power law bulge, a Miyamoto-Nagai disk, and an NFW dark
matter halo.  To allow for a range of galaxy masses and sizes, we also
wish to consider potentials with varying virial radius $R_{\rm vir}$ and virial
velocity $V_{H}$.  To do this, we consider a family of Milky-Way-like
potentials in which the mean density is the same, but mass and virial velocity
of the NFW halo 
vary with halo virial radius $R_{\rm vir}$ according to $M_{H} \propto R_{\rm vir}^{3}$ and
$V_{H}^{2} \propto GM_{H}/R_{\rm vir} \propto R_{\rm vir}^{2}$, with $V_H / \kms = R_{\rm vir}/\kpc $.  The
disk and bulge mass and size are rescaled in the same way.  

Within a given potential, we sample a few different footpoint
locations, and 
for each footpoint, we consider a range of the angular momentum $J$
(see \autoref{fig:streamline} and detailed parameter discussion in
\autoref{sec:param}).
Each footpoint location $R_0$ and choice of angular momentum $J/(\Omega R_0^2)$
defines a streamline.  For each streamline, we explore a
two-dimensional parameter space of the sonic point location
$z_{c}/R_{\rm vir}$ and Alfv\'en speed at the critical point $v_{A,c}$.  As discussed
in \autoref{sec:sonic}, this two-dimensional parameter space maps to
a two-dimensional parameter space of footpoint initial conditions 
for the wind at a distance $z=1\kpc$ above the disk midplane.  

The ISM in the coronal region may have a range of gas, magnetic, and
cosmic ray pressures. These depend on the midplane ISM properties as
well as the star formation activity, which drives a galactic fountain
that circulates gas from the midplane to coronal regions.  We
non-dimensionalize the problem so that the three relevant pressures
are captured as two ratios: the thermal gas pressure to cosmic ray
pressure ratio $P_0/\Pi_0=c_{s}^{2} (\Pi_0/\rho_0)^{-1}$, and the magnetic
field pressure to cosmic ray pressure ratio
$B_{0}^{2}/(8\pi\Pi_{0})=(1/2)v_{A,0}^2(\Pi_0/\rho_0)^{-1}$.
We note that $v_{A,0}^2$ and $\Pi_0/\rho_0$
are obtained from outputs of the wind
integration starting at the critical point and ending at the footpoint.
We also non-dimensionalize all of the velocities as ratios with respect to
$c_{s}$, which we set to be 10 km/s for a ``cool'' wind consisting of
warm-phase ISM gas.

We are interested in cases where the magnetic-to-cosmic ray pressure ratio
brackets equipartition by an order of magnitude (above and below).
Since this ratio is close to equipartition in the Solar neighborhood,
and the scale heights of these components are large, we expect that at
$z = 1$ kpc they remain roughly in equipartition.

\subsection{Sample Wind Solutions}\label{sec:sample}

\begin{figure*}
\includegraphics[width=\textwidth]{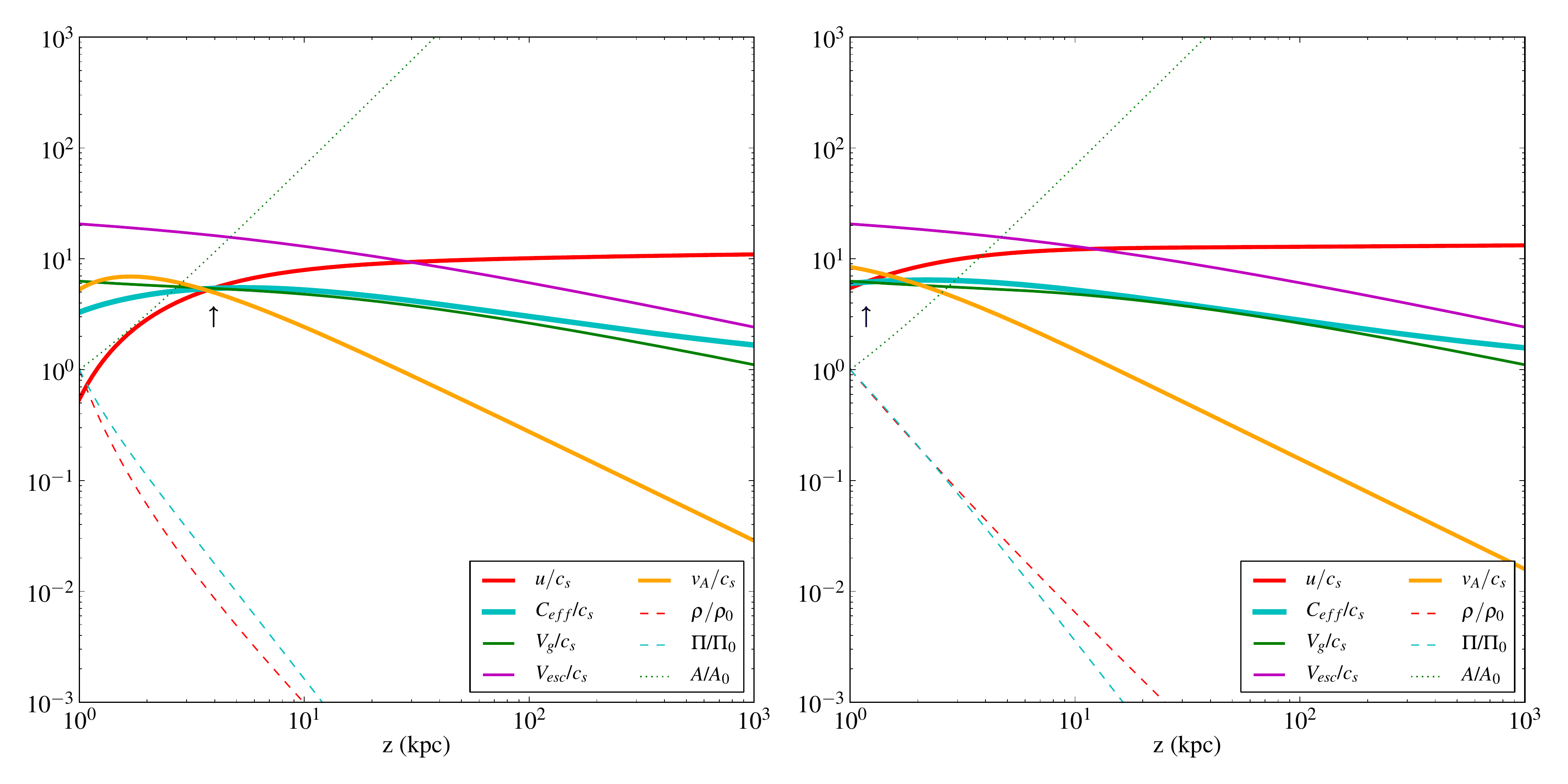}
\caption{ An example solution of the wind equation in a dwarf galaxy 
  potential with $V_{H} = 50$ km/s, launched at $R = 1$ kpc and $z = 1$
  kpc with no angular momentum. For the left and right panels, the wind
  is launched with initial velocity $u_{0} = 5\kms$ and $u_{0} = 50\kms$,
  respectively. In both panels, the footpoint cosmic ray and magnetic
  pressures are chosen to be equal, 
  $B_{0}^{2}/(8\pi\Pi_{0}) = 1$.  The black arrow indicates the critical
  point where $u=V_g=C_{\rm eff}$,
  equal to $54 \kms$ and $61 \kms$ for left and right panels.  
}\label{fig:windsol1}
\end{figure*}

\begin{figure*}
\includegraphics[width=\textwidth]{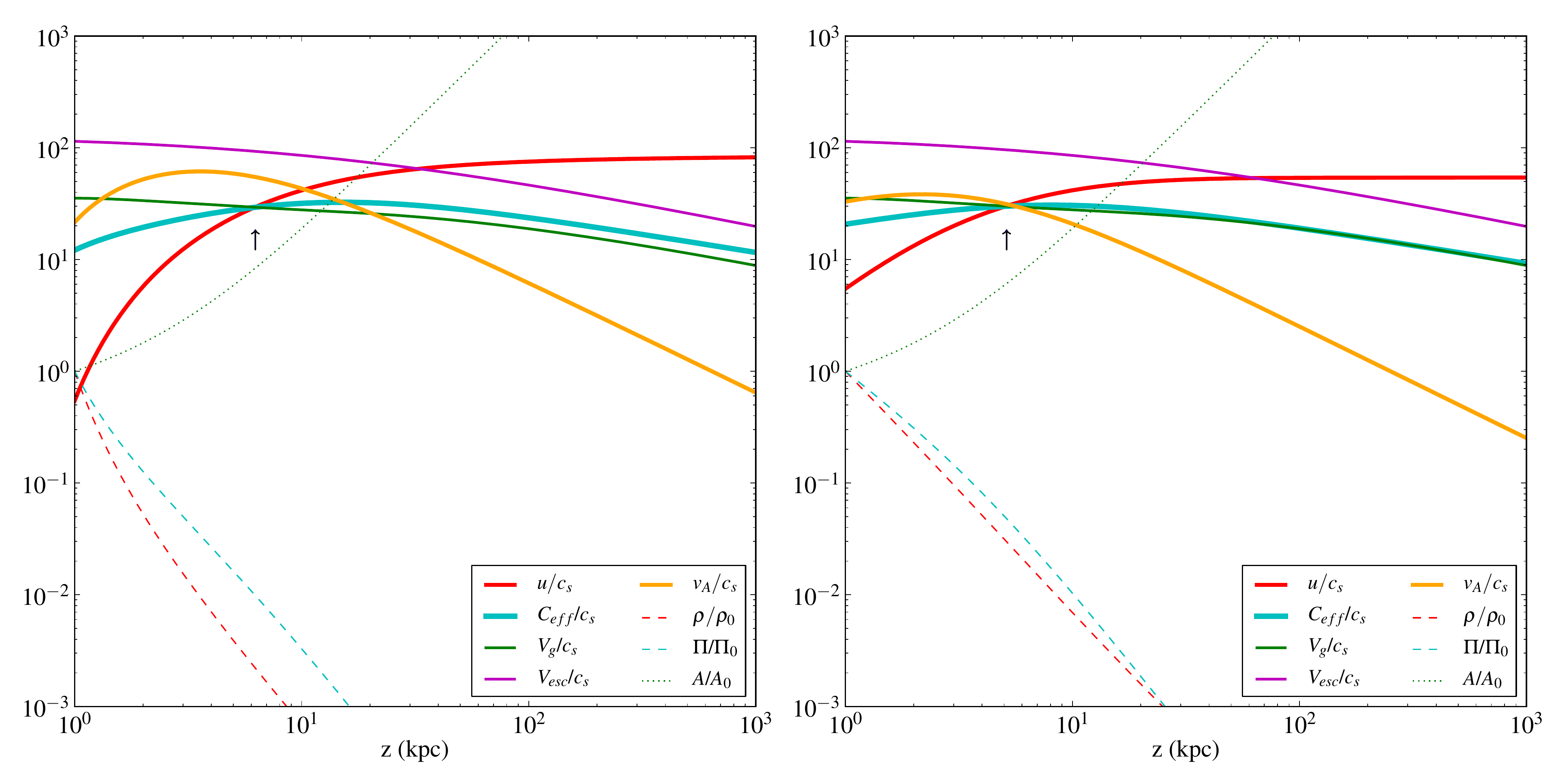}
\caption{
  A sample wind solution as in \autoref{fig:windsol1},
  except for a Milky Way potential with $V_{H} = 250$ km/s, and
  launched at $R = 4$ kpc.
  At the critical point, $V_{\rm g,c}= 290, 300 \kms$ for left and right panels,
  respectively.  
}\label{fig:windsol2}
\end{figure*}

Examples of wind solutions for a dwarf galaxy halo with $V_H=50\kms$
and a Milky Way-like halo with $V_H=250\kms$ are shown in
\autoref{fig:windsol1} and \autoref{fig:windsol2}.  For each halo
potential, cases with initial launch velocity $u_0=5\kms$ and
$u_0=50\kms$ are shown.  In all cases, the footpoint cosmic ray
pressure and magnetic pressure are chosen to be in equipartition.  For
the dwarf model, the footpoint radius is $R_0=1\kpc$, while for the
Milky Way model the footpoint radius is $R_0=4\kpc$.
The angular momentum parameter is set to $J=0$.
Specification
of $u_0$ and $B_0^2/(8\pi \Pi_0)$ selects a unique wind solution for
a given halo potential and streamline.

For all solutions shown, $u$ secularly increases with distance, while
$V_g$  secularly decreases.  $C_{\rm eff}$ increases outward inside the critical
point, and then decreases at large distance.  The Alfv\'en speed $v_A$ exceeds
$u$
inside the critical point, but drops off to
small values at large distance.  The density $\rho$ and cosmic ray pressure
$\Pi$ secularly decrease with distance.  In detail, $u$ becomes nearly
constant at large distance, which for a radial flow implies that
$\rho \propto (u A)^{-1} \propto r^{-2}$.  Thus, from \autoref{eq:vau},
$v_A \propto u \rho^{1/2} \propto r^{-1}$ at large distance, which in turn
implies that the effective sound speed declines slowly, as
$C_{\rm eff} \propto \rho^{(\gamma_{\rm cr}-1)/2}\propto \rho^{1/6}\propto r^{-1/3}$, 
at large distance (modulo flattening due to $c_{s}$).  
Since both $V_{g}$ and $C_{\rm eff}$ are equal at the sonic point and decrease slowly thereafter, they tend to be similar up until large distance where $C_{\rm eff} \sim c_{s}$. 
The escape speed $V_{\rm esc}\equiv \sqrt{-2(\Psi - \Psi_{\infty})}$ is larger than $V_H$ but decreases with distance, so that
$u$ eventually exceeds $V_{\rm esc}$ and in the absence of intervening
halo gas, the wind would escape.  In practice, wind propagation at large
distance would
ultimately be limited by interaction with surrounding halo gas.

\subsection{Wind Parameter Exploration}\label{sec:param}

\begin{figure*}
\includegraphics[width=\textwidth]{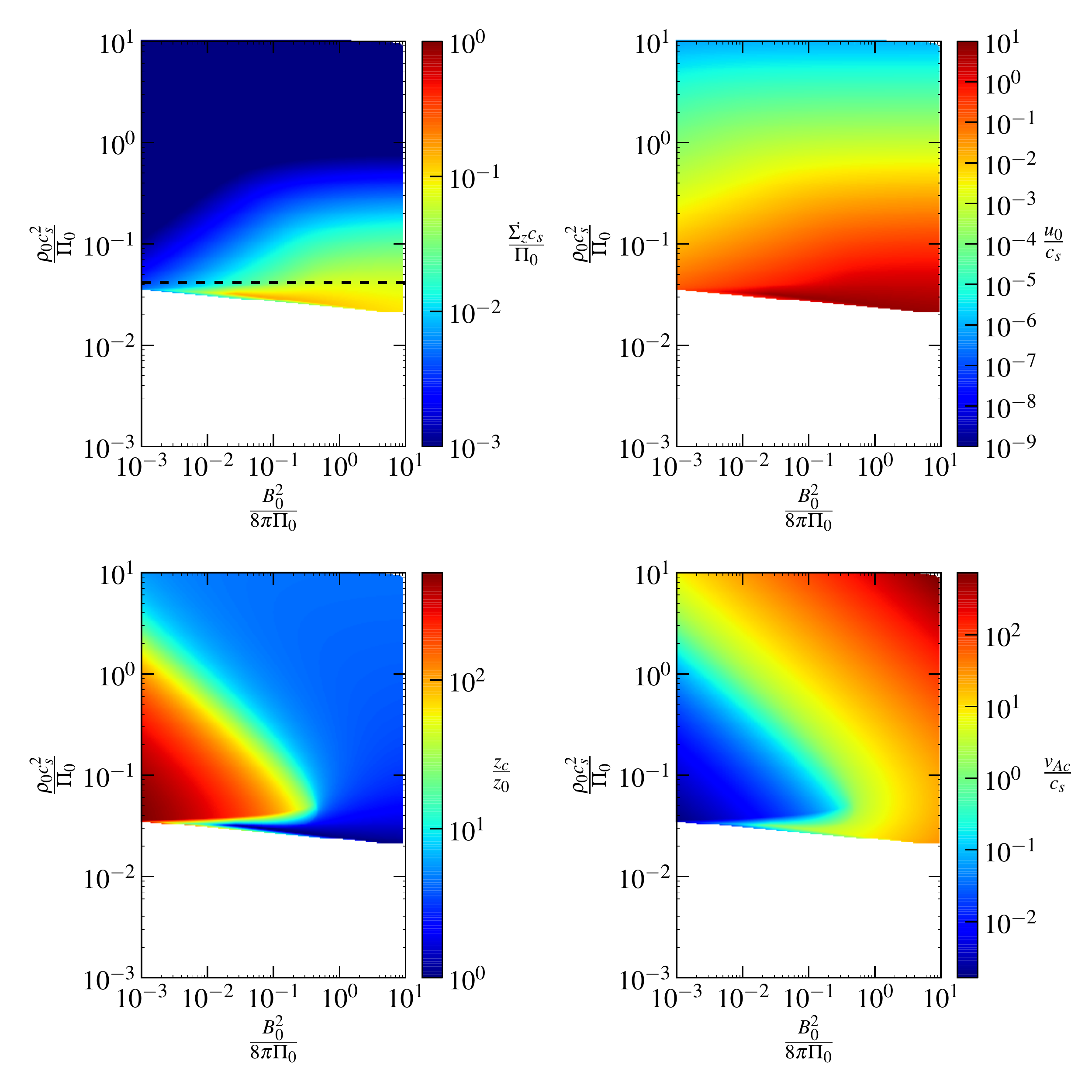}
\caption{ A two-dimensional parameter exploration of solutions of the
  wind equation in a dwarf galaxy potential with $V_{H} = 50$ km/s.
Results are shown for solutions on streamlines 
  launched at $R = 1.0$ kpc
  and $z = 1.0$ kpc with no angular momentum. A characteristic wind
  solution belonging to this set is shown in \autoref{fig:windsol1}.
  The top left panel shows in color scale results for
  the (dimensionless) mass-loss rate per unit area
  from the disk, $\dot \Sigma_z c_s/\Pi_0$, as a function 
  of thermal-to-cosmic-ray and magnetic-to-cosmic-ray pressure ratio
  at the streamline footpoint.
  The dashed horizontal line  is where $\Pi_0/\rho_0 = (V_H^2 - c_{s}^{2})$.  
  The top right panel shows the solution for $u_0$,
  the speed of the flow at the footpoint.  Bottom left and bottom right
  panels show solutions for
  the $z$ location and Alfv\'en speed at the critical point.
}\label{fig:panel1}
\end{figure*}

\begin{figure*}
\includegraphics[width=\textwidth]{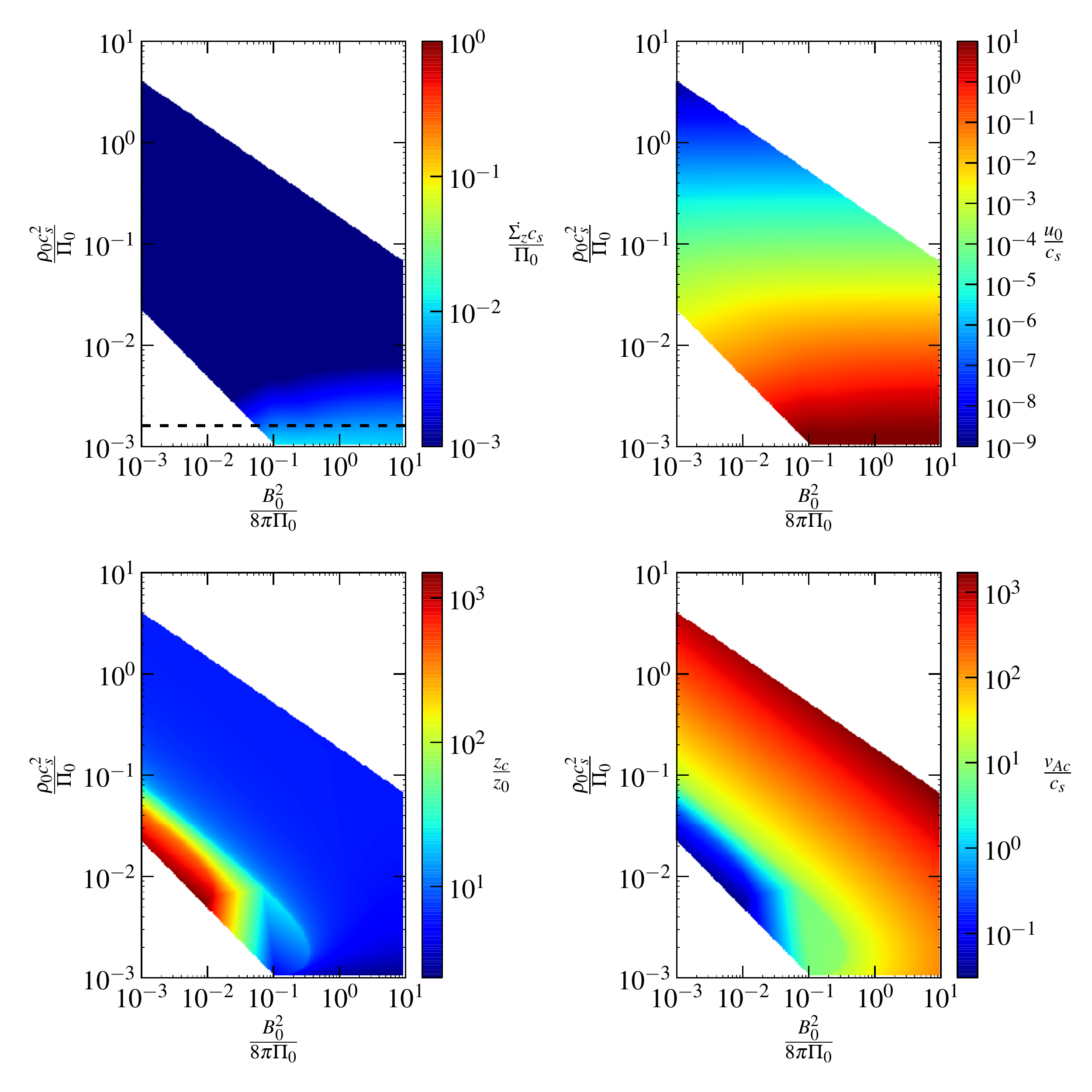}
\caption{
  Same as \autoref{fig:panel1}, except for a Milky Way-like galaxy potential
  with $V_{H} = 250$ km/s, launched at $R = 4$ kpc. A characteristic wind solution belonging to this set is shown in \autoref{fig:windsol2}.
}\label{fig:panel2}
\end{figure*}

\begin{figure*}
\includegraphics[width=\textwidth]{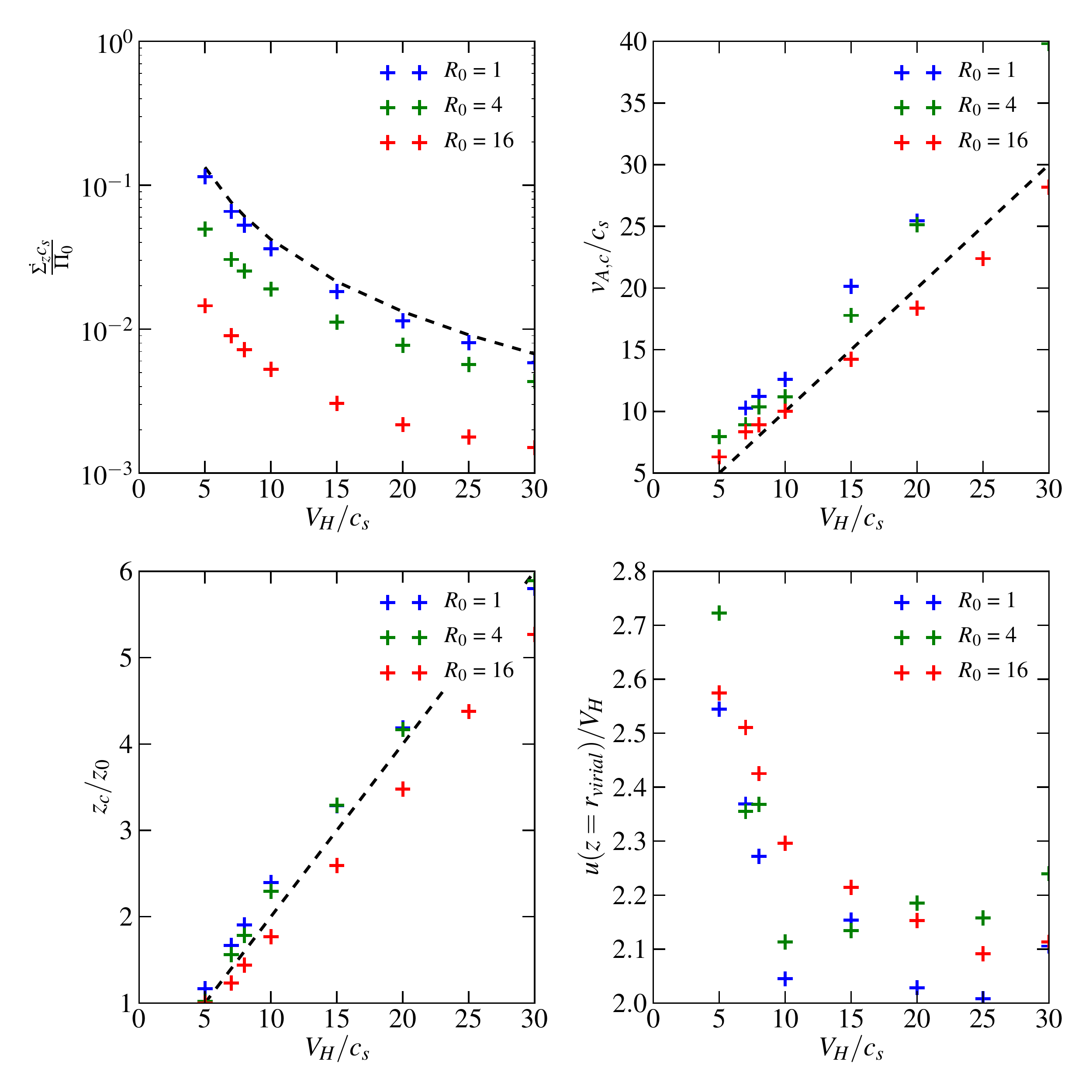}
\caption{ Results from wind solutions for a range of halo velocities
  $V_H$ and footpoint radii $R_0$.  For all cases, $\Pi_{0} = B_{0}^{2}/8\pi$
  and $u_{0} = 50\kms$ at the footpoint at height $z_0=1\kpc$, and $J=0$.  
  The top left panel shows that at constant $u_{0}$, the mass loss rate
  decreases with increasing $V_H$, and that footpoints originating at larger
  radii have lower (scaled) mass-loss rates.
  The $\dot{\Sigma}_{z}$ results are comparable to
  $\dot{\Sigma}$ from \autoref{eq:flux} setting $A_{c}/A_{0} = 1$
  (dashed black curve), with reduction at larger $R_0$
  in part due to lower $\hat s \cdot \hat z$.  The top
  right panel shows that the sonic Alfv\'en speed $v_{A,c}$ scales
  approximately linearly with $V_H$ (dashed black line represents
  $v_{A,c}=V_{H}$), and is slightly larger for smaller $R_0$. The bottom left
  panel shows that the location of the critical point is typically not
  far from the launch point for winds with near-equipartition cosmic
  ray and magnetic pressure, with $z_c \propto V_H \propto R_H$
  (dashed line).  Sonic points in more massive
  halos are further out because the wind must accelerate more to reach 
  $u = V_{g}$. The bottom right panel shows that by the time the wind
  reaches the virial radius, it would reach a speed a few times
 greater than $V_H$.
}\label{fig:plotmoon1}
\end{figure*}

We have extensively explored the parameter space of galaxies' potentials
and footpoint ISM properties.  In particular, we have considered
potentials with $V_H$ in the range
$50 - 300 \kms$.
Our standard set of footpoint locations is $R_0=1, 2, 4, 8, 16 \kpc$,
and we vary the angular momentum parameter $J$ by selecting values up to a
\revise{maximum} value for each footpoint, described in \autoref{sec:angmag}.

To explore a range of footpoint ISM conditions for each potential and
each streamline, in practice we begin by sampling a grid of critical point
locations and Alfv\'en speeds. Some of these
points yield footpoint solutions that
fall within a few orders of magnitude of equipartition between
gas, magnetic field, and cosmic ray pressure. Interpolating between
those points yields estimates for values of the critical point location
and Alfv\'en speed whose corresponding winds begin near desired
points in the space of footpoint pressure ratios. This allows us to fill in
the pressure space even though integration begins from the critical
point.

Every computation yields either a wind accelerating through the sonic point or
fails immediately by decelerating through the sonic point, which helps
to delimit the boundaries of the space in which interesting wind solutions
exist.  Here, we focus on wind solutions in which $u$ secularly increases
with distance.  Since accelerating winds require $C_{\rm eff,0} <
V_{g,0}$, a lower limit to $\rho_{0}c_{s}^{2}/\Pi_{0}$ is set by conditions
that yield $C_{\rm eff,0} = V_{g} \sim V_{H}$.  If $\rho_{0}/\Pi_{0}$ is too
low, $C_{\rm eff,0}$ exceeds $V_{g,0}$, and the wind does not accelerate. 
This lower limit is roughly illustrated by the black dashed horizontal line
denoting $\Pi_0/\rho_0 = (V_H^2 - c_{s}^{2})$
in the upper-left panels of \autoref{fig:panel1} and \autoref{fig:panel2}.

When $v_{A,c}/u_c$ is small, $C_{\rm eff}$ is decreasing through the critical point (see \autoref{eq:dsceff2}).
Since $V_{g}$ must decrease faster than $C_{\rm eff}$ for a critical transition
to exist, small $v_{A,c}$ ends up producing a sonic point
at a large distance. But to yield a sonic point at large distance,
$C_{\rm eff,0}$ must be large, and this implies small $\rho_{0}c_{s}^{2}/\Pi_{0}$.
Thus, as we not interested in solutions with sonic points at extremely
large distance, this places another lower limit on $\rho_{0}c_{s}^{2}/\Pi_{0}$. 
For example, the lower left sector of \autoref{fig:panel2} is excluded by these considerations, as can be seen by the large values of $z_c$ and the small values of $v_{A,c}$.

Winds with large $v_{A,c}$ tend to have strong acceleration, implying 
lower $u_{0}$ to reach a given $V_{g,c} \sim V_H$. Although solutions to
the wind equation exist for large $v_{A,c}$, we limit $v_{A,c}$ to avoid
unrealistically small  $u_{0}$. This consideration excludes the upper right sector of \autoref{fig:panel2}. 

For each wind solution, we are particularly interested in the mass-loss
rate.  Other parameters of interest are the critical point
location and Alfv\'en speed.  In addition, to decide whether a given cosmic-ray
driven wind solution can be realistically produced, it is important to consider
the footpoint velocity $u_0$.  Supernova-driven fountains can transfer
warm ISM gas from the midplane to the corona, but the velocity of
``fountain'' gas at $z\simgt \kpc$
distances above the midplane is typically $\simlt 100\kms$.\footnote{In
  particular, from self-consistent ISM/star formation/supernova feedback
  simulations, Kim \& Ostriker (2017, submitted) found that the mass
  of fountain gas (for Solar neighborhood conditions)
  exponentially decreases with velocity, with typical
  outflowing velocity $\sim 50\kms$
  at $z \sim 1-3\kpc$.}  A cosmic-ray driven wind must be able to match
its footpoint conditions
to the available gas mass and momentum flux into the corona from below, which
implies an upper limit on the value of $u_0$.

In characterizing the mass loss produced in our wind solutions,
we non-dimensionalize the mass flux by taking the ratio at the footpoint
to $\Pi_0/c_s$.
Considering only the $z$ component of the wind velocity to get the mass
loss per unit area of the galactic disk, we have 
\begin{equation}\label{eq:mdotz}
  \frac{\dot{\Sigma}_{z}c_{s}}{\Pi_{0}} = \frac{\rho_{0}u_{0,z}c_{s}}{\Pi_{0}}
  =\frac{\rho_{0}c_{s}^{2}}{\Pi_{0}}\frac{u_{0z}}{c_{s}}.
\end{equation}
Thus, for a given ratio of gas-to-cosmic-ray pressure at the footpoint,
the normalized
mass flux is set by the normalized vertical component of the
footpoint velocity, $u_{0,z}/c_s$.

Examples showing the space of two dimensional pressure ratios for
which wind solutions have
been found (on a given streamline in a given potential) are shown in
\autoref{fig:panel1} and \autoref{fig:panel2}.  For each point in
the identified wind solution space,
values of the mass-loss rate, the footpoint velocity, the vertical
distance to the critical point, and the Alfv\'en speed at the critical
point are shown in color scale in separate panels.  

From the top two panels in \autoref{fig:panel1} and
\autoref{fig:panel2}, the mass loss rate and initial wind velocity
appear to be primarily a function of the gas density with very little
dependence on the strength of the magnetic field, when magnetic and
cosmic ray pressures are within an order of magnitude of equipartition.
Furthermore, comparing
winds from massive galaxies (large $V_{H}$) to
dwarf galaxies shows that increasing $V_{H}$ shifts the solution space
towards lower density (lower $\rho_0 c_s^2/\Pi_0$), and also leads to
lower scaled mass loss (lower $\dot \Sigma_z c_s/\Pi_0$). This is
not qualitatively surprising, as the potential well is deeper (larger
$V_g \sim V_H$) in
a more massive galaxy, and therefore larger $C_{\rm eff}$ is needed to drive
outflows to reach escape speed.  Since $C_{\rm eff}^2 \sim \Pi/\rho$,
the mean density of winds in more massive galaxies must be lower if they
are to successfully escape.  In \autoref{sec:scaling}, we demonstrate 
analytically and numerically that a relationship is expected
between mass-loss rate and the ratio $\rho_0 c_s^2/\Pi_0$.  
Then, from the
definition of our dimensionless mass loss rate, a relationship between
the gas density and the mass loss rate also fixes the initial wind
velocity.

Naively, it might seem surprising that there is a lower bound on the
density (or an upper bound on the cosmic ray pressure) for which wind
solutions exist in dwarf  galaxies.  However, the reason for this
lower limit is that we are only interested in accelerating winds with
low initial velocity.   This requires $C_{\rm eff,0} < V_{g,0}$ at the footpoint, as
discussed in \autoref{sec:sonic}.  
Since $C_{\rm eff}^{2} \sim \Pi/\rho$, there is an upper limit on what
$C_{\rm eff,0}^{2}$ (and hence the cosmic ray pressure) can be that is still
consistent with a given (low) value of $V_g \sim V_H$. Lower
density winds with higher $C_{\rm eff,0}$ that are already escaping with
$u_{0} > C_{\rm eff,0} > V_{g,0}$ are mathematically allowed.
However, these are not of interest for the present work,
because they are not driven by cosmic ray pressure gradients above the
main body of the ISM.

Some general characteristics of winds are illustrated in
\autoref{fig:plotmoon1}.  In this figure, we consider a range of halo
velocities ($V_H=50-300\kms$) and footpoint radii ($R_0=1, 4,
16\kpc$).  We show results of solutions for which the footpoint
magnetic field is in equipartition with the cosmic ray pressure
($B_0^2/(8\pi)=\Pi_0$) and the footpoint launch speed is $u_0=50\kms$,
with angular momentum $J=0$ and footpoint height $z_0=1\kpc$.  For
each wind solution, we show the scaled mass-loss rate
($\dot \Sigma_z c_s/\Pi_0$, \autoref{eq:mdotz}), the Alfv\'en speed
at the critical point ($v_{A,c}/c_s$), the vertical distance of the critical point
from the footpoint ($z_c/z_0$), and the flow velocity at large distance
relative to the halo velocity ($u[z=R_{\rm vir}]/v_H$). As might be expected,
$v_{A,c}$ and the wind velocity at the virial radius are
roughly proportional to $V_H$, and $z_c$ increases roughly linearly
with $R_{\rm vir}$.  The critical point is relatively near the launch point
when $B_0^2/(8\pi)=\Pi_0$, as is also evident in \autoref{fig:panel1} and
\autoref{fig:panel2}.  For a given $V_H$, $\dot \Sigma_z c_s/\Pi_0$ is
larger for smaller footpoint radius $R_0$, and the differential effect
is largest at small $V_H$.  The dependence of $\dot \Sigma_z c_s/\Pi_0$
on $R_0$ is largely because
streamlines are most vertical for small $R_0$.  
The scaled mass-loss rate $\dot \Sigma_z c_s/\Pi_0$ decreases
with $V_H$; we discuss the specific scaling behavior
(dashed curve) in \autoref{sec:scaling}.

\subsection{Wind Scaling Relations}\label{sec:scaling}

\autoref{eq:dsceff2} shows that $C_{\rm eff}$ increases outward provided
$v_A/u >0.64$, which implies that 
$v_{A}/u \gtrsim 1$, and in practice $v_A/u\gg 1$
(see \autoref{fig:windsol1}, \autoref{fig:windsol2}) 
for most of the evolution between the footpoint and the critical point.

In the limit of $v_{A} \gg u$,
\autoref{eq:pi} becomes 
\begin{equation}
\frac{\Pi}{\Pi_{0}} \approx \left(\frac{v_{A}}{v_{A0}}\frac{A}{A_0}\right) ^{-\gamma_{\rm cr}} = \left(\frac{\rho}{\rho_{0}}\right)^{\gamma_{\rm cr}/2}
\end{equation}
and the effective sound speed (see \autoref{eq:ceff2} and \autoref{eq:dpidrho})
becomes
\begin{equation}\label{eq:ceffrho}
  C_{\rm eff}^{2} \approx \frac{\gamma_{\rm cr}}{2}\frac{\Pi_{0}}{\rho_{0}}
  \left(\frac{\rho}{\rho_{0}}\right)^{\gamma_{\rm cr}/2-1}.
\end{equation}
As discussed in \autoref{sec:bernoullieq}, this implies that 
for $\gamma_{\rm cr} = 4/3$, $C_{\rm eff}^{2} \propto \rho^{-1/3}$, which increases
outward as $\rho$ decreases outward.

\begin{figure*}
\includegraphics[width=\columnwidth]{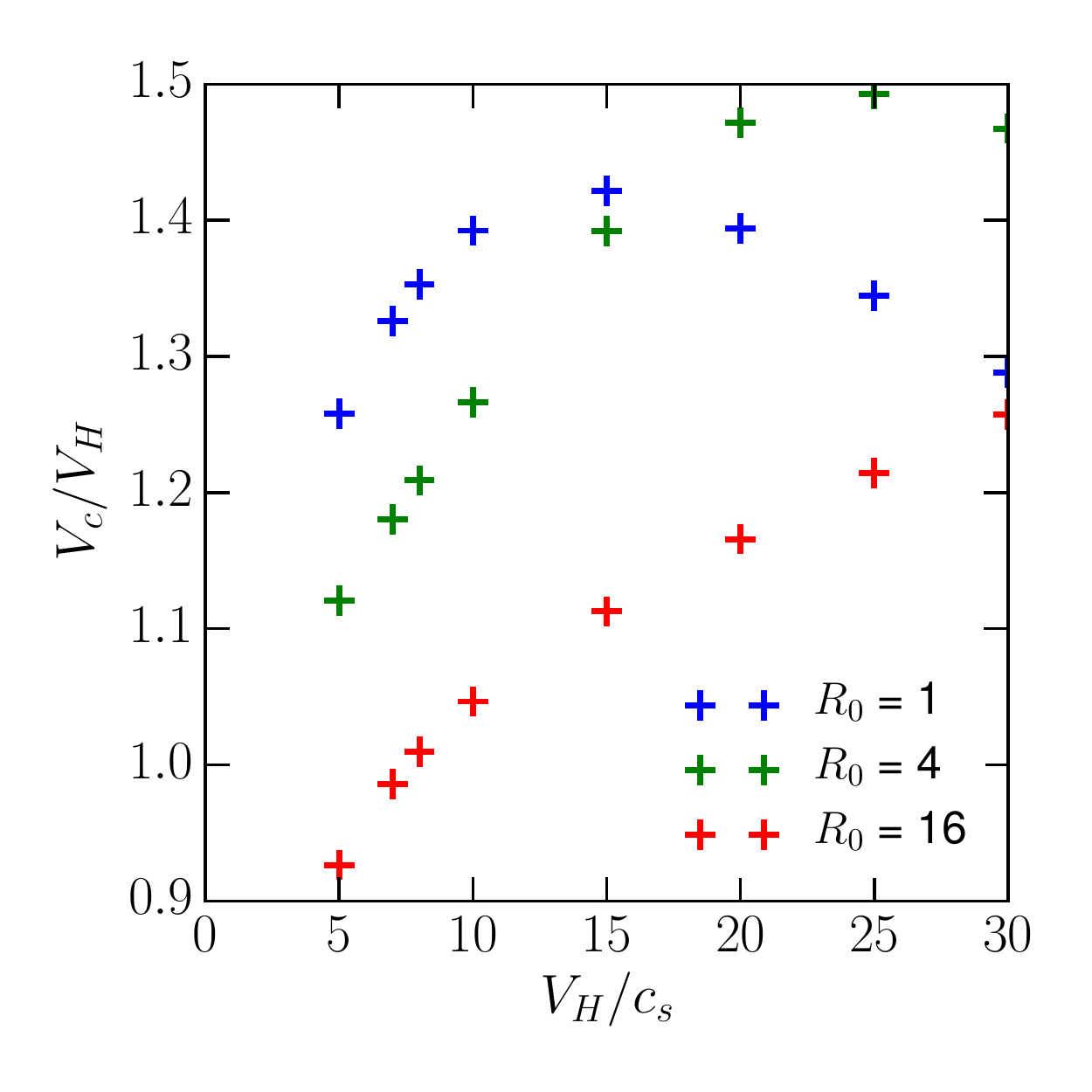}
\caption{
The critical velocity $V_c\equiv u_{c} = V_{g,c} = C_{\rm eff,c}$ as a function of halo velocity. Parameters are as in \autoref{fig:plotmoon1}.
}\label{fig:plotvcvh}
\end{figure*}

At the critical point, $C_{\rm eff,c}^{2} = V_{g,c}^{2} =
u_{c}^{2}$. Furthermore, since a typical galactic rotation curve is
close to flat, $V_{g,c} \sim V_{H}$ where $V_{H}$ is a characteristic halo
velocity.  In particular, \autoref{fig:plotvcvh} shows that the range of ratios $V_{g,c}/V_{H} = 0.9-1.5$ for $V_{H} = 50-300\kms$.
Finally, conservation of mass flux implies
$\rho_0 u_0 A_0 =\rho_c u_cA_c $.
Combining these relations (and using $\gamma_{\rm cr}=4/3$),
\autoref{eq:ceffrho} may be solved for the footpoint 
mass flux ratio or pressure ratio as
\begin{equation}\label{eq:scaleflux}
\frac{\dot \Sigma c_s}{\Pi_0}\equiv 
  \frac{u_0  \rho_0c_s}{\Pi_0} \sim
\left(\frac{2}{3}\right)^3 
\left(\frac{\rho_0 c_s^2}{\Pi_0}\right)^{-2}
  \left(\frac{V_H}{c_s}\right)^{-5}  \frac{A_{c}}{A_{0}}
\end{equation}  
or
\begin{equation}\label{eq:scaling}
  \frac{\Pi_{0}}{\rho_{0} c_s^2} \sim \frac{3}{2}
  \left(\frac{V_{H}}{c_s}\right)^{5/3}
  \left(\frac{u_0}{c_s}\right)^{1/3}
  \left(\frac{A_{0}}{A_{c}}\right)^{1/3} .
\end{equation}

\autoref{eq:scaleflux} shows that for a fixed halo potential ($V_H$), at
large $\rho_0 c_s^2/\Pi_0$ the normalized mass flux $\dot \Sigma
c_s/\Pi_0$ and footpoint velocity $u_0/c_s$ must be small.  This is
consistent with the behavior evident in the numerical wind solution
results shown in the top panels of \autoref{fig:panel1} and
\autoref{fig:panel2} for $V_H=50\kms$ and $V_H=250\kms$, respectively.
Also, since $u_0 < V_{g,c} \sim V_H$ and $A_0 <A_C$, from \autoref{eq:scaling}
a lower limit on the footpoint density is given by
$\rho_0 c_s^2 /\Pi_0 \sim (2/3) (V_H/c_s)^{-2}$. This limit is roughly shown with a dashed horizontal line in \autoref{fig:panel1} and \autoref{fig:panel2}.

\begin{figure*}
\includegraphics[width=\textwidth]{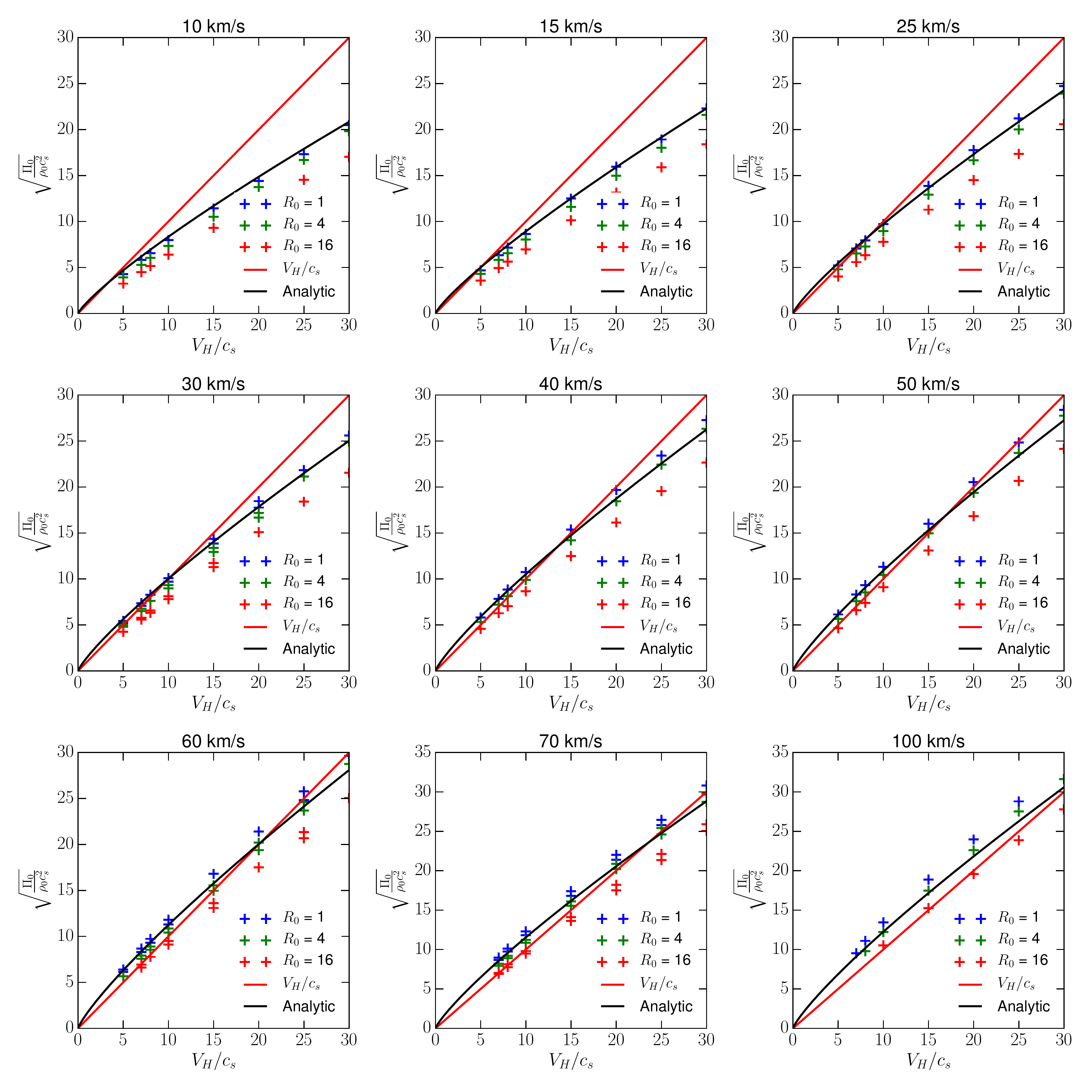}
\caption{
  Plots of $\sqrt{\Pi_{0}/\rho_{0}c_{s}^{2}}$ from
  wind solutions with a range
  of halo potential velocities $V_H$ and footpoint radii $R_0$.
  Other launch conditions are as in \autoref{fig:plotmoon1}, except we
  show results for a range of launch velocities $u_0=10-100\kms$
  (labeled at top of each panel).  In each panel, the red line shows
  the naive order-of-magnitude estimate ($ \Pi_0/\rho_0 \sim V_{H}^2 $), while
  the black curve labeled ``Analytic'' shows \autoref{eq:scaling}, setting
  $A_{c}/A_{0} = 1$.  
}\label{fig:plotmoon2}
\end{figure*}

\begin{figure*}
\includegraphics[width=\textwidth]{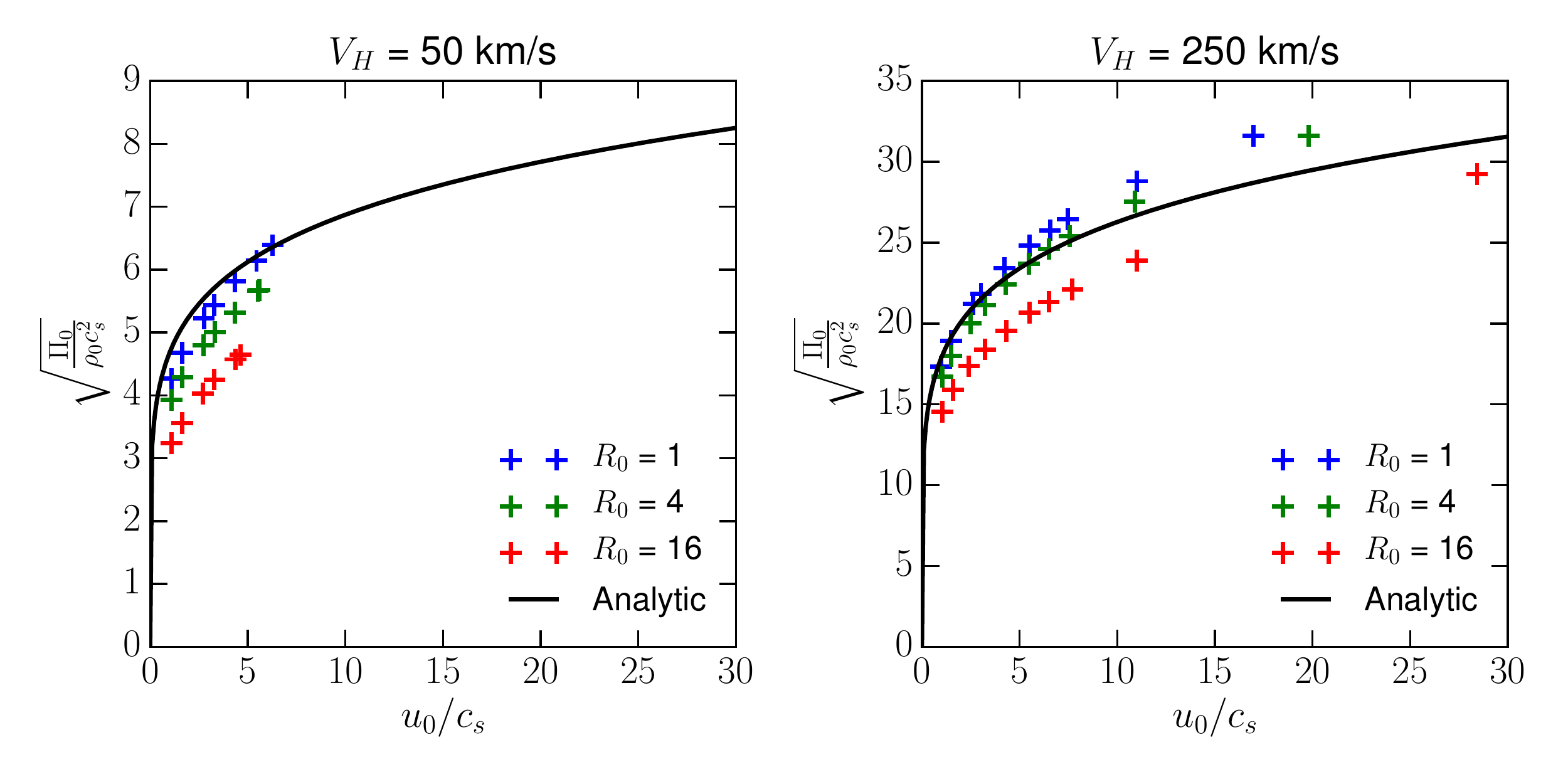}
\caption{
Results for $\sqrt{\Pi_{0}/\rho_{0}c_{s}^{2}}$ 
vs. $u_{0}/c_{s}$, holding $V_H$ fixed.  Parameters are otherwise
as in \autoref{fig:plotmoon2}.
The left and right panels show results for a dwarf and massive
halo ($V_H=50\kms$ and $250\kms$), respectively.  The 
black curve labeled ``Analytic'' shows the prediction of
\autoref{eq:scaling} with $A_{c}/A_{0} = 1$.
}\label{fig:plotmoon3}
\end{figure*}

The scaling relation in \autoref{eq:scaling} can be compared to the dependence
of the footpoint pressure ratio $\Pi_0/(\rho_0 c_s^2)$  on 
$V_H$ and $u_0$ found in our numerical wind solutions.
\autoref{fig:plotmoon2} shows the dependence
of $\Pi_0/(\rho_0 c_s^2)$ on $V_{H}$  for
actual solutions of the wind equation with a series of $u_{0}$ values,
compared to the analytic estimate \autoref{eq:scaling}
taking $A_c/A_0=1$.  Evidently,
the analytic prediction is in quite good agreement with the numerical
results.  \autoref{fig:plotmoon2} also shows that the solutions are
insensitive to the footpoint radius.  Dimensional analysis would
suggest that if the momentum flux associated with the cosmic ray footpoint
pressure, $\Pi_0$, is directly transferred to momentum flux in a wind with
characteristic velocity $\sim V_H$ and density $\sim \rho_0$, then
one would naively expect $\Pi_0/\rho_0 \sim V_H^2$.  
Red lines in each panel of \autoref{fig:plotmoon2} shows that this naive
expectation is not bad as a zeroth order estimate, but that it increasingly
fails to fit the true wind solutions at high $V_H$ and low $u_0$.  
Instead, the prediction $\Pi_0/\rho_0 \sim (3/2)V_H^{5/3} u_0^{1/3}$ of
\autoref{eq:scaling} fits the numerical results well over the full parameter
space. Similarly, in \autoref{fig:plotmoon3} we compare the results of wind
integrations to the predicted dependence of the footpoint pressure ratio on
$u_0$, again showing good agreement.  

\autoref{eq:scaling} can be rearranged to provide an estimate for
the ``carrying capacity'' mass flux in a galactic disk wind that
originates in a coronal region where the cosmic ray pressure is
$\Pi_0$ and ISM material at $T\sim 10^4\Kel$
is fed from below by a supernova-driven fountain
flow with velocity $u_0$.  This carrying capacity is
\begin{equation}\label{eq:flux}
\dot \Sigma =\rho_0 u_0 \sim \frac{2}{3} \Pi_0 V_H^{-5/3} u_0^{2/3} \left(\frac{A_c}{A_0}\right)^{1/3}.
\end{equation}
Of course, $u_0<V_{g,c}\sim V_H$, so $\dot \Sigma \simlt \Pi_0/V_H$
for low velocity halos.  We compare the carrying capacity to
$\dot{\Sigma}_{z}$ in the top left panel of
\autoref{fig:plotmoon1}. The difference is at most $10\%$ for $R_{0} =
1\kpc$, a factor of 2 for $R_{0} = 4\kpc$, and a factor of 7 for
$R_{0} = 16\kpc$. The variation for different $R_{0}$ is primarily due
to the geometric factor $\hat s \cdot \hat z$. For smaller $R_{0}$,
the streamline following the gravitational potential starting at $z =
1\kpc$ is more vertical, whereas distant $R_{0}$ have more radial
streamlines.

The relation in \autoref{eq:flux} shows that winds driven by cosmic
ray pressure are not expected to follow either the ``momentum'' ($\dot
\Sigma_z \propto V_H^{-1}$) or ``energy'' ($\dot \Sigma_z \propto
V_H^{-2}$) scalings that have commonly been adopted in ``subgrid''
wind models in galaxy formation simulations \citep{Somerville2015}.
Instead, the scaling with $V_H$ is intermediate between these two
limits, and an additional dependence on the ``feeding'' velocity $u_0$
is also present. We emphasize that the far-field
wind velocity {\it does}, however, scale nearly
linearly with $V_H$, as shown in \autoref{fig:plotmoon1}.

Finally, we remark that \autoref{eq:flux} 
is the carrying capacity for winds driven by cosmic ray pressure, but
more generally for any driving effective pressure $P_{\rm e}$,
\autoref{eq:winds} will still
hold for $C_{\rm eff}^2 \rightarrow dP_{\rm e}/d\rho$ (see \autoref{eq:ceff2alt}),
and $C_{\rm eff,c}^2 = u_c^2 =V_{g,c}^2 $ must still hold at the critical point.  
If $P_{\rm e} \propto \rho^{\gamma_e}$, then 
for a galactic wind with $V_{\rm g,c}\sim V_H$
the generalization of \autoref{eq:flux} is
\begin{equation}\label{eq:fluxgen}
\dot \Sigma =\rho_0 u_0 \sim \gamma_e P_{\rm e,0} V_H^{-(\gamma_e+ 1)} u_0^{\gamma_e} \left(\frac{A_c}{A_0}\right)^{1-\gamma_e}
\end{equation}
where $P_{e,0}$ is the driving pressure at the footpoint.
With $C_{\rm eff}^2 = \gamma_e (P_{\rm e,0}/\rho_0)(\rho/\rho_0)^{\gamma_e-1}$,
$0<\gamma_e <1$ is required for $C_{\rm eff}$ to increase
with distance such that a steady, accelerating wind
is able to make a critical transition in a $V_g \sim V_H =const.$
galactic potential.
\footnote{\revise{Note that we find $V_{g} \sim V_{H}$ for disk winds with footpoints and critical points at scales $\sim 1-10$ kpc.
    For galactic center quasi-spherical winds with critical points at smaller scales ($\sim 200$ pc), \cite{2016ApJ...819...29B,2017ApJ...835...72B} find that the halo component can be neglected. In this case $V_{g} \sim V_{H}$ does not hold
    so \autoref{eq:fluxgen} does not apply and $\gamma < 1$ is not required for a wind.}}
Cosmic-ray driven winds have
$\gamma_e\approx\gamma_{\rm cr}/2 =2/3$ (inside the critical point).
\autoref{eq:fluxgen} shows that
any simple pressure-driven galactic \revise{disk} wind will have
dependence on $V_H$ between the ``momentum-driven''and ``energy-driven'' 
scalings,
i.e. $\propto V_H^{-(1+\gamma_e)}$ with $1<1+\gamma_e <2$.  In contrast to case of
a galactic \revise{disk} wind with an extended potential, a wind from a point
mass (or any truncated mass distribution) has $V_g$ decreasing outward
$\propto r^{-1/2}$, 
so that a steady wind with a critical transition may have
$C_{\rm eff}$ also decrease outward, compatible with 
$\gamma_e \ge 1$.  This is a key distinction
between pressure-driven Parker-type winds \revise{(which would include
  quasi-spherical galactic center winds for which the halo potential is
  unimportant)}  and galactic \revise{disk} winds 
(see \autoref{fig:cartoon}). 

\subsection{Angular momentum and magnetic field dependence}\label{sec:angmag}

\begin{figure*}
\includegraphics[width=\textwidth]{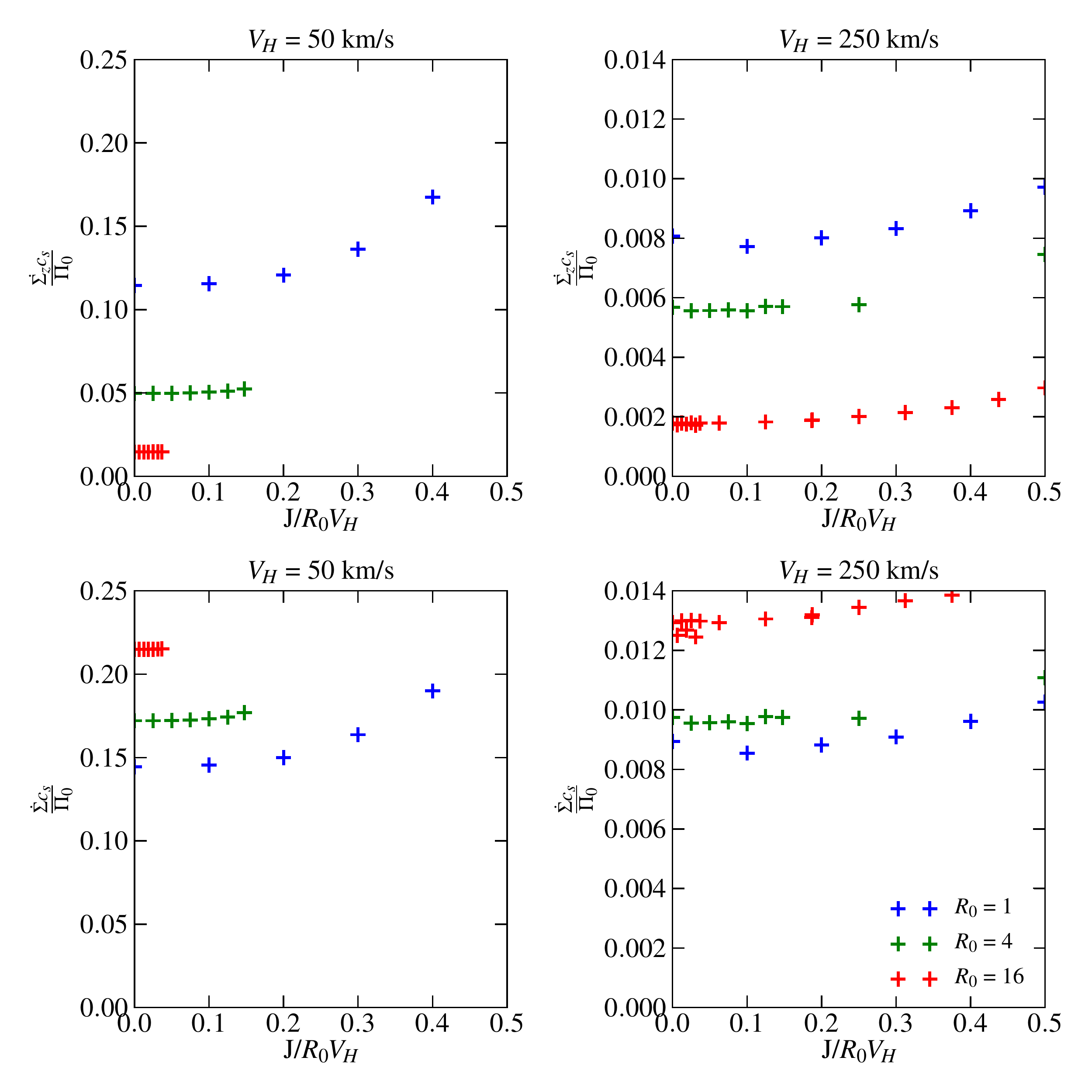}
\caption{ Dependence of streamline flux $\dot \Sigma = \rho_0 u_0$ and
  disk mass-loss rate per unit area
  $\dot \Sigma_z = \hat s \cdot \hat z \dot \Sigma$
  on the angular momentum parameter
  $J$.  Top: $\dot{\Sigma}_{z}c_{s}/\Pi_{0}$ vs. $J/(R_{0}V_{H})$.
  Bottom: $\dot{\Sigma}c_{s}/\Pi_{0}$ vs. $J/(R_{0}V_{H})$.
  Left panels show results from a dwarf galaxy potential
  ($V_H=50\kms$) and right panels
  show results from a Milky Way-like potential ($V_H=250\kms$).
For all cases, $\Pi_{0} = B_{0}^{2}/8\pi$
and $u_{0} = 50\kms$ at the footpoint at height $z_0=1\kpc$.
Points with different colors correspond to streamlines with
footpoint radii $R_0=1, 4, 16\kpc$.  The mass-loss rate increases slightly
with $J$, but overall the effect of rotation is modest. 
  }\label{fig:plotangj}
\end{figure*}

Angular momentum of the flow has a small effect on the wind. The centrifugal
force produces acceleration in the rotating frame in the $\hat{R}$
direction, and as shown in \autoref{sec:hydro} this effect can be incorporated
in an effective potential $\Psi$.  The centrifugal force partly compensates
for the inward force of gravity, which near the disk is primarily in
the $-\hat R$ direction.  Since 
we assume streamlines follow the gradient
of the effective potential, and angular momentum reduces the gradient of
$\Psi$ in
the $\hat{R}$ direction, the resulting streamlines are more vertical
at higher $J$.  This effect is shown in \autoref{fig:streamline}. 

Since
angular momentum opposes inward gravitational acceleration, it
decreases $V_{g}$ along the streamline. We do not explore large
angular momentum $J > 0.5R_{0}V_{H}$ because the effective potential
produces a gradient that would be unrealistic for streamlines, 
turning around towards $R = 0$ at large $z$. 

For a nearly vertical streamline with large $J$, at $z \gg R$ the
gravitational and centrifugal components of the effective potential
gradient (which is related to streamline direction $\hat{s} =
\nabla\Psi/|\Psi{}|$ by assumption), respectively drop off as
$\hat{z}/z$ and $-\hat{R} R_{0}^{2}/R^{3}$. Since $R$ remains roughly
constant and $z$ is increasing, the centrifugal term eventually
dominates the streamline.  This leads to a streamline which
unrealistically turns towards $R = 0$ at large $z$ if $J > J_{\rm max}$.
We numerically determine the maximum value $J_{\rm max}$ for each value of
$R_{0}$ in \autoref{fig:streamline} and note that typically
$J_{\rm max} > 0.5R_{0}V_{H}$. Hence, we avoid those values. \revise{This
  consideration determines
  the range of streamlines depicted in \autoref{fig:streamline}.}

For given mass flux $\dot \Sigma = \rho_0 u_0$ along streamlines,
the mass-loss rate per unit area in the disk
$\dot \Sigma_z$ is lower by a factor
$\hat s \cdot \hat z = [(dR/dz)^{2} + 1]^{-1/2}$.  More vertical
  streamlines, with smaller $dR/dz$, therefore have a larger $\dot \Sigma_z$,
  other things being equal.  
By examining \autoref{fig:streamline}, this effect is
small for small footpoint radii $R_{0}$, since the fractional change
in $\hat s \cdot \hat z$ is small for varying $J$.

\autoref{fig:plotangj} shows results for mass-loss rates in
two different halo potentials, at a range of footpoint locations, for
varying angular momentum parameter $J$.  
The top panels show that the mass-loss rate per unit area in the disk
depends more strongly on $R_0$ (and corresponding streamline geometry)
than on the angular momentum $J$.   The bottom panels show that
larger $R_{0}$ cases correspond to larger $\dot{\Sigma}$
(because $V_{g}$ is slightly smaller at the critical point; see \autoref{sec:scaling}).
In comparison, the top panels show that the geometric effect
is strong enough to reverse this trend for
$\dot{\Sigma}_{z} = \hat s \cdot \hat z\dot{\Sigma}$, with larger $R_0$
yielding smaller $\dot \Sigma_z$.
Note that increasing $J$ decreases the upper limit
on $u_{0}$ for which there is an accelerating solution. For example, at large
values of $J/R_{0}V_{H}$ and fixed $u_{0} = 50\kms$, accelerating solutions
exist for high $V_{H}$ halos but not low $V_{H}$ halos,
as evident in \autoref{fig:plotangj}. 

\revise{
In this work, we have 
ignored the toroidal component of the magnetic field and any associated magnetic stresses. Work by \cite{1996A&A...311..113Z} includes these magnetic forces in a rotating galaxy, finding that increasing the magnetic field strength by a factor of 3 leads to roughly 1.4 - 2 times more mass loss.
If we included magnetic forces, they would provide an additional acceleration that could increase $u^{2}$ up to $u_{\phi}^{2}/(1 - M_{A}^{2})$, where $M_{A} = u/v_{A}$. Both $M_{A}$ and $u_{\phi}$ are small inside the critical point for the winds we study 
so the acceleration from magnetic pressure forces would be small. 
}

\begin{figure*}
\includegraphics[width=\textwidth]{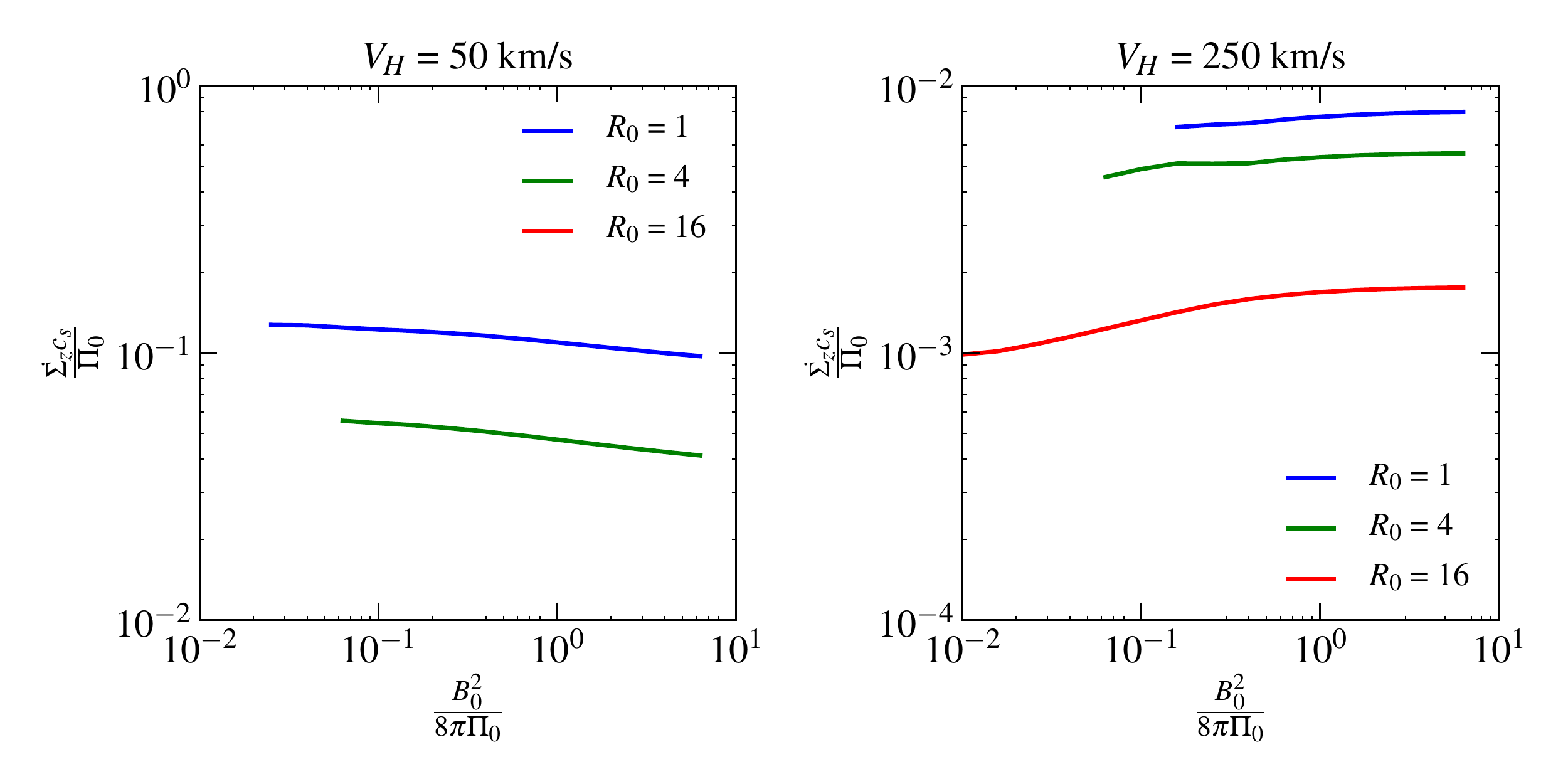}
\caption{Mass-loss rate per unit area $\dot{\Sigma}_{z}c_{s}/\Pi_{0}$
  vs. magnetic pressure $B_{0}^{2}/8\pi\Pi_{0}$ for $u_{0} = 50$ km/s in a dwarf
  galaxy potential with $V_{H} = 50\kms$ and a Milky-Way like potential with
  $V_H=250\kms$. Winds originate from $z_{0} = 1$ kpc and $R_{0}
  = 1,4,16$ kpc with no angular momentum ($J = 0$). The mass loss rate
  varies only very weakly with magnetic field strength.
}\label{fig:plotmag}
\end{figure*}

Since we do not 
include magnetic forces, the magnetic field only affects
winds through the value of the Alfv\'en speed $v_{A}$ \revise{(associated with
the poloidal field component)}, which controls the
streaming rate of cosmic rays. This in turn affects the evolution of 
$C_{\rm eff}$, which must increase relative to $V_{g}$ to produce a critical point where $C_{\rm eff} = V_{g}$. To have $C_{\rm eff}$ increase outward, $v_{A}/u > 0.64$ is required
(see \autoref{eq:dsceff2}).
Since $v_{A}$ relative to $u$ only determines
the effective adiabatic index of the cosmic ray fluid, $v_A$ does not
directly appear in the scaling relation \autoref{eq:scaling}
(for sufficiently large $v_{A}$), and therefore 
the wind \revise{is expected to depend} only weakly on the strength of the magnetic field
$B$.
\revise{We find that in wind solutions }
the magnetic field strength at the base of the flow ($B_{0}$) indeed has
a relatively small effect on the wind properties.
This is evident in \autoref{fig:plotmag}, in which changing the
magnetic pressure by three orders of magnitude leads to less than order unity
change in the mass loss rate.  This is also evident in the top left
panel of \autoref{fig:panel1} and \autoref{fig:panel2}. At smaller
magnetic field strengths, increasing $B_{0}$ leads to increased mass
loss since a larger $v_{A}$ allows a larger $u_{0}$ under the
constraint that $v_{A}/u$ must be large enough to produce an
accelerating wind with a sonic transition.

\subsection{Implications for Mass Loading of Galactic Winds}\label{sec:mdot}

Mass fluxes for our wind solutions are all given in units of $\Pi_0/c_s$, with
values in the range $\sim 0.001 -0.1$ in these units  (see
\autoref{fig:panel1}, \autoref{fig:panel2}, \autoref{fig:plotmoon1},
\autoref{fig:plotangj}, \autoref{fig:plotmag}).  The physical value of
the mass flux therefore depends on the cosmic ray pressure (or energy density)
in the region where the wind originates.  Consider as an example the
Solar neighborhood, where the local cosmic ray pressure is
$P_{\rm cr} \sim 0.6 {\rm eV}\pcc$ \citep{2015ARA&A..53..199G}.
Using $c_s =10\kms$, the dimensional
factor for the mass-loss rate would be $\Pi_0/c_s \rightarrow P_{\rm cr}/10\kms
\sim 0.15 \sfrunit$.  For
$\dot \Sigma_z c_s/\Pi_0 \sim 0.004$, as might be appropriate for the
Solar neighborhood with $u_0=50\kms$ (see \autoref{fig:plotmoon1}) ,
the result is $\dot \Sigma_z\sim 5 \times 10^{-4}\sfrunit$.  The
corresponding 
footpoint number density of the wind at $z=1\kpc$
would be $n_0=8 \times 10^{-4} \pcc$ (assuming mean molecular weight of 1.4 $\mh$).
This mass-loss rate is $\sim 20\% $ of
the observed star formation rate estimated in the Solar neighborhood,
$2.5\times 10^{-3}\sfrunit$ \citep{2009AJ....137..266F}.  

More generally, we showed that the ``carrying capacity''
estimate in \autoref{eq:flux} follows the
numerical results quite well, especially for small $R_0$
(see \autoref{fig:plotmoon1}),
so it is useful to rewrite it in dimensional form \revise{(with $A_c/A_0=1$)}as
\begin{equation}\label{eq:fluxdim}
  \dot \Sigma \sim 3 \times 10^{-3} \sfrunit
\left(\frac{\Pi_0}{1 {\rm eV \pcc}} \right)
  \left(\frac{V_{H}}{200 \mathrm{km s^{-1}}}\right)^{-5/3}
\left(\frac{u_{0}}{50 \mathrm{km s^{-1}}}\right)^{2/3}.
\end{equation}
Lower halo velocity $V_H$ or higher feeding velocity $u_0$ increases the
mass-loss rate.
\second{
  The corresponding density of hydrogen nuclei
  in the wind at the footpoint in the launching
region (above the main ISM disk) is 
  \begin{equation}\label{eq:ndim}
  n_0 \sim 1.8 \times 10^{-3} \pcc
\left(\frac{\Pi_0}{1 {\rm eV \pcc}} \right)
  \left(\frac{V_{H}}{200 \mathrm{km s^{-1}}}\right)^{-5/3}
\left(\frac{u_{0}}{50 \mathrm{km s^{-1}}}\right)^{-1/3}.
\end{equation}
Note that this density is much lower than the typical midplane 
density of both the cold and warm ISM, but based on
numerical simulations (e.g. Kim \& Ostriker 2017, submitted) is 
similar to mean densities of warm ``fountain'' gas in galactic
disk corona regions.  
}

Mass-loss in galactic winds is often characterized in terms of the ``mass
loading,'' defined as the ratio of the local wind mass-loss rate to the local
star formation rate, $\beta \equiv \dot \Sigma_{\rm wind}/\Sigma_{\rm SFR}$, where
$\dot \Sigma_{\rm wind} = \dot \Sigma_z$ in the present notation.
Because cosmic rays are produced in the supernova remnants associated with
explosions from young, massive stars, the cosmic ray pressure at
the disk midplane
\second{likely scales with the}
star formation rate,
$P_{\rm cr} = \eta_{\rm cr} \Sigma_{\rm SFR}$.  For the Solar neighborhood,
$\eta_{\rm cr} \sim 600 \kms$.  Other components of the midplane
pressure, including the thermal pressure and turbulent kinetic and
magnetic pressures, \second{are expected} to be proportional to $\Sigma_{\rm SFR}$
\citep{2010ApJ...721..975O,Ostriker2011}
with respective ``feedback yield'' coefficients $\eta_{\rm th}$,
$\eta_{\rm turb}$, $\eta_{\delta B}$, etc., that can be computed with
detailed numerical simulations of the ISM including star formation
and feedback \citep{2011ApJ...743...25K,Kim2013,Kim2015b,2016arXiv161203918K},
such that the total pressure is $P_{\rm tot}=\eta_{\rm tot} \Sigma_{\rm SFR}$.

\second{Assuming $\Pi_0$ is comparable to the midplane cosmic ray pressure, we then have for the}
predicted mass-loading factor for cosmic-ray driven winds,
\begin{align}
  \frac{\dot \Sigma_{\rm wind, cr}}{\Sigma_{\rm SFR}}&=
  \frac{\eta_{\rm cr}}{c_s} \frac{\dot \Sigma_z c_s}{\Pi_0}
\label{eq:massload1}
  \\
  &\sim 0.8   \left(\frac{\eta_{\rm cr}}{600 \kms}\right)
\left(\frac{V_{H}}{200 \mathrm{km s^{-1}}}\right)^{-5/3}
\left(\frac{u_{0}}{50 \mathrm{km s^{-1}}}\right)^{2/3}
\left(\frac{A_c}{A_0}\right)^{1/3} \hat z \cdot \hat s.
\label{eq:massload2}
\end{align}
In applying \autoref{eq:massload1},
numerical results for $\dot\Sigma_z c_s/\Pi_0$ can be drawn from the figures,
while
\autoref{eq:massload2} comes from \autoref{eq:flux}.
\second{Assuming} $\eta_{\rm cr}/c_s \sim 100$, the mass-loading factor for
cosmic-ray driven winds will exceed unity when
$\dot{\Sigma}_{z}c_{s}/\Pi_{0} \simgt 0.01$.  From the top-left
panels of \autoref{fig:panel1} and \autoref{fig:panel2}, the mass
loading is order unity or higher near equipartition
($B_{0}^{2}/8\pi\Pi_{0} \sim 1$) for sufficiently low $\rho_{0}$, which corresponds
to high $u_0$ (top-right panels of \autoref{fig:panel1} and
\autoref{fig:panel2}).  
From \autoref{eq:massload2}, mass-loading for cosmic-ray driven winds
is expected to exceed unity in dwarf galaxies where $V_H \simlt 200\kms$,
provided $u_0 \sim 50\kms$ is consistent with galactic fountain
flows that carry gas into the corona (see e.g. Kim \& Ostriker 2017, submitted).

\second{
Finally, we emphasize that Equations \ref{eq:flux}, \ref{eq:fluxdim}, and \ref{eq:massload1} represent carrying capacities, and hence are upper limits for the mass flux or mass-loading of a cosmic-ray driven warm-gas wind that originates in disk corona regions and is fed by a galactic fountain from below.  Of course, the wind mass-loss rate cannot exceed the mass feeding rate from below.  While the general dependence of the fountain mass flux on local disk parameters is not presently known, current numerical MHD simulations of supernova-driven outflows do show a mass-loading factor of the warm fountain near unity at height of a few times the warm-ISM scale height (Kim \& Ostriker 2017, submitted; see also \cite{2016martizzi} and \cite{2017lbo}).}

\second{
The asymptotic specific energy of the gaseous wind is $(1/2) V_{\infty}^2$, where $V_{\infty}/V_H \sim 2$ from
\autoref{fig:plotmoon1}.  This implies that the asymptotic energy loading of the wind (defined as ratio of wind energy to energy injected by supernovae) is then $\sim (V_H/500 \kms)^2$ times the mass-loading factor, where we have assumed $100 M_\odot$ in stars are formed for every $10^{51}\erg$ of energy injected by supernovae.  Using the fiducial $\eta_{\rm cr}$ and
$u_0$ in \autoref{eq:massload2}, this yields an energy loading less than $\sim 10\%$ for $V_H < 300\kms$, as must be the case if the wind ultimately derives its power from cosmic rays that are accelerated in supernova
remnants.  However, we caution that $\eta_{\rm cr}$ in \autoref{eq:massload2} need not be a constant, and it is not known how it may depend on local ISM properties.  The energy flux in cosmic rays at the base of the wind is $(u_0 + v_{A,0})3 \Pi_0  \hat z \cdot \hat s$.
If we assume that this is of order 10\% of the energy input rate from supernovae,
$(700 \kms)^2\Sigma_{\rm SFR}$, this places a practical upper limit on the product $(u_0 + v_{A,0})\eta_{\rm cr} \hat z \cdot \hat s$.
}
\section{Summary and Discussion}\label{sec:sum}

In this paper, we have used one-dimensional (1D) steady-state models to explore
the properties of galactic disk winds driven by cosmic ray pressure.  In
contrast to previous studies of cosmic-ray driven disk winds
using steady-state 1D idealizations \citep{Ipavich1975,Breit1991},
we adopt a streamline shape that is specifically motivated by ``downhill''
flow in a realistic galactic effective potential (including bulge, disk, 
halo, and a centrifugal term).
Also, as our main interest is in understanding how large quantities of
relatively cold gas may be accelerated to escape from a deep potential
well, we adopt an isothermal equation of state with $c_s=10\kms$
($T \sim  10^{4}$ K) for which 
thermal pressure forces are negligible and cosmic ray pressure forces
provide the needed acceleration, rather
that considering hot outflows (as from galactic center starburst regions)
that are driven by both thermal and cosmic ray pressure
\citep[e.g.][]{Everett2008}.

A key feature of winds driven by cosmic ray pressure is that the
square of the effective sound speed $C_{\rm eff}^2 = d\Pi/d\rho$
increases $\propto \rho^{-1/3}$ with decreasing $\rho$ when $v_{A}/u$
is sufficiently large (see \autoref{eq:dsceff2} and
\autoref{sec:scaling}),
and generally increases relative to the
squared gravitational velocity $V_{g}^{2} = d_{s}\Psi/d_{s}\ln{A}$
inside the critical point (see \autoref{sec:ceff2}).  In contrast, an
adiabatic thermal wind cools as it expands and $\rho$ drops, so that
the sound speed strictly decreases outward as $C_{\rm eff}^2 \propto
\rho^{\gamma_e-1}$ for $\gamma_e >1$.  Thermal-pressure driven
galactic \second{disk} winds face an inherent challenge, as $V_{g}$  must decrease
faster than $C_{\rm eff}$ in order to make a steady sonic transition,
but a galactic potential including an extended dark matter halo has a
nearly flat rotation curve with $V_g^2\sim V_H^2$ out to large
radii. For cosmic rays, $C_{\rm eff}^2$ increases outward because
streaming at the Alfv\'en speed implies $\Pi \propto n_{\rm cr}^{\gamma_{\rm cr}}
\propto (v_A
A)^{-\gamma_{\rm cr}}\propto \rho^{\gamma_{\rm cr}/2}\propto
\rho^{2/3}$.  \autoref{fig:cartoon} shows the characteristic
differences between galactic winds driven by cosmic ray pressure and
classical Parker stellar winds, while \autoref{fig:windsol1} and
\autoref{fig:windsol2} show examples of our full numerical solutions.

We performed a wide parameter exploration over
halo virial velocities $V_H= 50 - 300 \kms$
(masses $M_H \sim 10^{10} - 10^{12} M_{\odot}$), 
streamline footpoint radii $R_0= 1 - 16 \kpc$,
angular momentum $J= 0 -0.5 R_{0}V_{H}$, 
ratios of footpoint magnetic pressure to cosmic ray pressure
$B_0^2/(8\pi \Pi_0) =10^{-3}-10$,
and ratios of footpoint
gas pressure to cosmic ray pressure
$\rho_0 c_s^2/\Pi_0= 10^{-3}-10$.

Our steady wind solutions have the following properties:

\begin{enumerate}

\item Winds are sub-Alfv\'enic ($u<v_A$) at least until reaching the
  critical point
  (see \autoref{sec:ceff2}).  After making a
  sonic transition, where $C_{\rm eff,c}=u_c=V_{g,c}\sim V_H$,
  acceleration slows and $u$ flattens out,
  while $v_A$ declines rapidly and $C_{\rm eff}$ declines slowly
  at large distance.

\item
For fixed $u_0$,
  the dimensionless mass-loss rate   $\rho_0 u_{0,z}/\Pi_0$
  is insensitive to the footpoint
  magnetic-to-cosmic-ray pressure ratio $B_0^2/(8\pi \Pi_0)$ and
  angular momentum (\autoref{fig:plotangj}, \autoref{fig:plotmag}).
The value of $\rho_0 u_{0,z}/\Pi_0$ 
  increases at low gas-to-cosmic-ray pressure ratio $\rho_0 c_s^2/\Pi_0$
(\autoref{fig:panel1}, \autoref{fig:panel2}).  
  However, the solution space for steady, accelerating
  winds to exist at all has a minimum 
  footpoint gas density, corresponding to $\Pi_0/V_H^2 \simlt  \rho_0$.

\item For $B_0^2/(8\pi \Pi_0)=1$ and footpoint velocity
  $u_0=50\kms$, over the full range
  of $V_H$ we find that the mass-loss rate 
  $\dot \Sigma_z=\rho_0 u_{0,z} \sim 0.01-0.1 ~\Pi_0$
  (decreasing at larger $V_H$ and increasing slightly with $R_0$),
  the critical point is close to the disk
  $z_c \sim 1-5\kpc$ (increasing linearly with $V_H$) with
  $v_{A,c}\sim V_H$, and at the virial radius $u$ is 2 - 3 times $V_H$
  (\autoref{fig:plotmoon1}).  
\end{enumerate}

We show that our numerical integration
results are in good agreement with a simple
analytic prediction relating footpoint properties of ``successful''
steady wind solutions with
the halo velocity as $\Pi_0/(\rho_0 u_0^{1/3})\sim (3/2) V_H^{5/3}$,
as given in \autoref{eq:scaling}. This can be recast as the
carrying capacity of a wind driven by cosmic ray pressure, with
streamline mass flux $\dot \Sigma \sim (2/3) \Pi_0 V_H^{-5/3} u_0^{2/3}$
(\autoref{eq:flux} or \autoref{eq:fluxdim}).  The footpoint velocity
$u_0$ that enters the mass-loss estimate
is presumably limited by the supernova-driven
fountain flow that carries gas from the midplane to the ``coronal''
region above the disk.  For galaxies with potentials similar
to the Milky Way, \autoref{eq:massload2}
suggests that the mass-loss rates for winds driven by cosmic ray pressure
will be only slightly lower than the star formation rates.
Mass loss
could significantly exceed star formation for dwarf galaxies.

An interesting feature of cosmic ray driven winds is their dependence
on the halo velocity. Whereas nominally the wind mass loading
$\beta =\dot \Sigma_{\rm wind}/\dot \Sigma_{\rm SFR} \propto V_{H}^{-1}$
for ``momentum driven'' winds and $\beta \propto V_{H}^{-2}$ for ``energy driven''
winds \citep{Murray2005,Somerville2015}, 
\autoref{eq:massload2} argues that $\beta \propto V_{H}^{-5/3}$
for \second{galactic disk winds} driven by cosmic ray pressure. This  power law
is in between the ``momentum'' and ``energy'' scalings,
is intriguingly similar to that in observations by \cite{Chisholm2017},
and is also consistent with
other observations (see \autoref{sec:intro}).  We remark that more
generally, steady galactic \revise{disk}
winds driven by any gamma-law pressure force
would have $\beta \propto V_H^{-(\gamma_e +1)}$ for $0<\gamma_e <1$.

Our work has several limitations. For example, we do not include
cosmic ray diffusion,
and we do not include effects of
magnetic pressure or tension forces on the flow. 
We also do not model the winds from a full disk but rather individual
non-interacting streamlines. A full disk would have non-uniform
structure and a distribution of cosmic ray pressures, gas densities,
and launching velocities from gas motions. Our model is unable to
incorporate possible effects of interaction between streamlines.
Furthermore, we treat the gas as a single-phase medium, but in reality
the warm medium in galactic disk coronal regions at $z\simgt \kpc$ would
have a volume filling factor below unity, with ``warm fountain'' gas
intermixed with hot gas (e.g. Kim \& Ostriker 2017, submitted).
The effects of volume filling factor on mass loss are uncertain, especially
as cosmic ray pressure forces on the gas are mediated by the interaction
of both the cosmic rays and gas with magnetic fields.  
To move beyond these limitations will require
full numerical MHD simulations of a multiphase ISM, including
self-consistent star formation and feedback, with a cosmic ray treatment
that includes streaming at the Alfv\'en speed along magnetic field lines.

While our models are idealized in many respects, our results provide
evidence that cosmic-ray driven winds may be quite important to the
evolution of galaxies, especially at $V_H\simlt 200 \kms$.  Our
analysis makes clear the distinctive physics behind cosmic-ray driven
winds, also providing scaling relations that may prove useful for
tests of and comparisons to
fully three-dimensional numerical implementations.  With the
possibility that cosmic ray pressure may drive more mass out of dwarf
galaxies than is locked up in stars, there is strong motivation to
include a realistic treatment of cosmic rays
in future galaxy formation simulations.

\section*{Acknowledgments}
We are grateful to the referee, Ellen Zweibel, for an insightful report, and Eliot Quataert for helpful suggestions.  
This work was supported by the National Science
Foundation under grant AST-1312006 and NASA under grant
NNX17AG26G to ECO, and grant
DGE-1148900 providing a Graduate Research Fellowship to SAM.

\appendix

\section{Ion Neutral Damping}\label{sec:ind}

To estimate the effect of ion-neutral damping we compare the streaming instability growth rate with the ion neutral damping rate. The growth rate is \citep{1969ApJ...156..445K}:
\begin{equation}
  \Gamma_{\rm CR} \sim \Omega_{0} \frac{n_{\rm CR}}{n_{\rm i}}
  \left(  \frac{v_{\rm D}}{v_{\rm A}} - 1\right)
\end{equation}
for ion cyclotron frequency $\Omega_{0}$, cosmic ray number number
density $n_{\rm CR}$, ion number density $n_{i}$ \second{(corresponding to
  mass density $\rho=\mu n_i)$},
and mean drift velocity of the cosmic ray distribution
$v_{\rm D}$.  We write $f_{D} = \left(  (v_{\rm D}/v_{\rm A}) - 1\right)$.

We note that
\begin{equation}
  \frac{n_{\rm CR}
  }{n_{\rm i}}
  \sim \frac{\Pi}{\rho{}c^{2}}
\end{equation}
and use \autoref{eq:scaling} so that at the base of the wind 
\begin{equation}
  \Gamma_{\rm CR} \sim  
\Omega_{0}
  \left(\frac{V_{H}}{c}\right)^{5/3}
  \left(\frac{u_0}{c}\right)^{1/3}
  f_{D}.
\end{equation}

The damping rate 
is \citep{1969ApJ...156..445K}:
\begin{equation}
  \Gamma_{\rm in} =\frac{1}{2} n_{n} \langle\sigma{}v\rangle
\end{equation}
for neutral number density $n_{n}$ and rate coefficient
$\langle\sigma{}v\rangle$, 
\second{where we assume a mostly-ionized medium}.
From \cite{1971ApL.....8..189K},
$\langle \sigma v \rangle =$ 1.53 to 8.40 $\times 10^{-9} \cm^3 {\rm s}^{-1}$
for $T=100$ to $10^4$ K.

Setting $\Gamma_{\rm CR} > \Gamma_{\rm in}$ as the condition for
ion neutral damping to be ignored, this requires
\begin{equation}
  n_{n}f_{D}^{-1} \lesssim 3 \pcc
  \left(\frac{B}{\mu{}{\rm G}}\right)
    \left(\frac{V_{H}}{200 \kms}\right)^{5/3}
    \left(\frac{u_0}{50 \kms}\right)^{1/3}
    \left(\frac{\langle\sigma{}v\rangle}{10^{-9}\cm^{3}{\rm s}^{-1}}\right)^{-1}
\end{equation}
This says that ion-neutral damping may be neglected provided that
$n_n/f_D$ is not too large.  If $n_n$ is small, that  means that $f_D$ can
also be very small (i.e. $v_D \rightarrow v_A$); larger $n_n$ would require
larger drift.  
A lower estimate,
taking $f_D\sim 1$, $V_{H} = 50 \kms$, and a rate coefficient of $10^{-8}\cm^{3}{\rm s}^{-1}$, gives $n_{n} \lesssim 0.03 \pcc$.
This is easily satisfied for the parameter regime we consider,
since even the ion density is only $\sim 10^{-3} \cm^{-3}$
for $\dot \Sigma_z \sim 10^{-3} \Msun \kpc^{-2} \yr^{-1}$
\second{(see also \autoref{eq:ndim} more generally)}.
The mass-loss rate would have to be very high, and the neutral fraction
very large, for ion-neutral damping to be significant.

\second{Finally, we note that for primarily-neutral gas in  higher density
  clouds, ion-neutral collisional damping is much stronger and cosmic rays
  are therefore expected to stream rapidly through such clouds, whether
  within the ISM or in galactic winds 
  \citep{2011ApJ...739...60E}.
}  

\section{Effective Sound Speed}\label{sec:ceff2}

By combining \autoref{eq:rho}, \autoref{eq:pi}, \autoref{eq:va},
\autoref{eq:vau}, and \autoref{eq:ceff2}, the effective sound speed
can be written purely as a function of $\rho$ and the ratio $v_{A}/u$.  Then,
using $d(u/v_A)/d\ln\rho = -(1/2)u/v_A$, we have
\begin{equation}
 \frac{1}{(C_{\rm eff}^2 -c_s^2)} \frac{dC_{\rm eff}^2}{d\ln\rho}=
  \frac{\gamma_{\rm cr}\left(1 + \frac{1}{2}\frac{v_A}{u}\right)^2 -
    1 - \frac{7}{4}\frac{ v_A}{u} -
    \frac{1}{2}\frac{v_A^2}{u^2}}{(1+\frac{v_A}{u})(1+\frac{v_A}{2u})}.
\end{equation}
Since $C_{\rm eff}^2 -c_s^2$ is positive and $d_s\ln\rho<0$
($\rho$ decreases outward), 
$C_{\rm eff}$ will increase outward ($d_sC_{\rm eff}^2 >0$)
provided that the sign of the right-hand
side is negative.  For $\gamma_{\rm cr}=4/3$, this is true for $v_A/u>0.64$.

The linear dependence of $\rho^{-1}$ on $u$ and $A$ also allows us to
simplify our treatment of the sonic transition
(\autoref{sec:sonic}) from \autoref{eq:lhopital}. That is,
$(d_{s}A)\partial_{A} =
(Ad_{s}\ln{A})\partial_{A} = 
(Ad_{s}\ln{A})(\partial_{A}\rho^{-1})\partial_{\rho^{-1}}
= -(d_{s}\ln{A})\partial_{\ln\rho}$. Similarly,
$\partial_{u} =
(\rho^{-1}/u)\partial_{\rho^{-1}}
= -(1/u)\partial_{\ln\rho}$. 
The partial derivatives with respect to $u$ assume holding $A$ constant and vice versa.
Solving \autoref{eq:lhopital} for $d_{s}u$, the behavior of the wind at the critical
transition is then given by a quadratic

\begin{equation}\label{eq:quadratic}
\begin{split}
  0 &= ((-1/u)\partial_{\ln\rho}C_{\rm eff} - 1)(d_{s}u)^{2}
  - 2\partial_{\ln\rho}C_{\rm eff}d_{s}\ln{A}(d_{s}u)
  - ud_{s}\ln{A}(\partial_{\ln\rho}C_{\rm eff}d_{s}\ln{A} -d_{s}V_{g} )
  \\
  &\equiv a(d_{s}u)^{2} + b (d_{s}u) + c
\end{split}  
\end{equation}
with a solution 
\begin{equation}
d_{s}u = \frac{b}{-2a} + \frac{\sqrt{b^{2}-4ac}}{-2a}.
\end{equation}
Since $C_{\rm eff}$ changes slowly, $a < 0$. Hence, there is an accelerating wind passing through the sonic point whenever $b > 0$ so that $d_{s}u > 0$. This corresponds to when $\partial_{\rho^{-1}}C_{\rm eff} > 0$. However, it is also possible to attain $d_{s}u > 0$ when $b < 0$, as long as $-4ac > 0$, so that the determinant is larger than $b$. This corresponds to $-\partial_{\ln\rho}C_{\rm eff}d_{s}\ln{A} > d_{s}V_{g}$. This is simply a mathematical demonstration of the qualitative property that the wind begins with $u_{0} < C_{\rm eff,0} < V_{g,0}$ and evolves so that eventually $V_{g} < C_{\rm eff} < u$. In order for $C_{\rm eff}$ and $V_{g}$ to change order, $C_{\rm eff}$ must increase relative to $V_{g}$. This concept is roughly illustrated in \autoref{fig:cartoon}. For a typical galactic potential with a nearly flat rotation curve, $V_{g}$ slightly decreases and is nearly constant. Thus, it is sufficient for $-\partial_{\ln\rho}C_{\rm eff} > 0$, so at the critical transition point, it is necessary for $v_{A} \gtrsim 0.64u$. Before this point, since $v_{A}/u$ is strictly decreasing, $v_{A} \gtrsim u$ throughout the evolution of an accelerating wind with a smooth sonic transition.   
 
Another possible family of accelerating solutions to \autoref{eq:quadratic} under the assumption $a < 0$ is 
\begin{equation}
d_{s}u = \frac{b}{-2a} - \frac{\sqrt{b^{2}-4ac}}{-2a}. 
\end{equation}
This requires $b>0$ so that $-\partial_{\ln\rho}C_{\rm eff} > 0$ and hence $d_{s}C_{\rm eff} > 0$, and simultaneously $-4ac < 0$, so that $d_{s}V_{g} > -\partial_{\ln\rho}C_{\rm eff}(d_{s}\ln{A}) > 0$. Again, since $b > 0$, this leads to $v_{A} \gtrsim u$. For such sonic point conditions, two branches of solutions are possible, but this second branch of solutions only occurs for gravitational potentials where $V_{g}$ is increasing.

\bibliographystyle{aasjournal}
\bibliography{refs}{}

\end{document}